\documentclass[
reprint,
onecolumn,
superscriptaddress,
dvipsnames,
notitlepage,
amsmath,amssymb,
aps
]{revtex4-2}
\usepackage[left=1.5cm,right=1.5cm,top=2cm,bottom=2cm]{geometry}
\usepackage{graphicx}% Include figure files
\usepackage{dcolumn}% Align table columns on decimal point
\usepackage{bm}% bold math
\usepackage{hyperref}% add hypertext capabilities
\usepackage[mathlines]{lineno}% Enable numbering of text and display math
\usepackage{multirow}
\usepackage{url}
\usepackage{float}
\usepackage{comment}
\usepackage{ragged2e}
\usepackage{adjustbox}
\usepackage{setspace}
\usepackage{adjustbox}

\usepackage[normalem]{ulem}
\usepackage{xcolor}
\usepackage{orcidlink}
\usepackage{flushend}

\usepackage{graphicx}% Include figure files
\usepackage{dcolumn}% Align table columns on decimal point
\usepackage{bm}% bold math
\usepackage{hyperref}% add hypertext capabilities
\usepackage{color}
\usepackage{multirow}

\hypersetup{pdfstartview=FitV,colorlinks=true,linkcolor=blue,citecolor=red,filecolor=black,urlcolor=blue}

\pdfoutput=1
\usepackage[T1]{fontenc}
\usepackage{amsmath,amssymb}
\usepackage{epsfig}
\usepackage{subfigure}
\usepackage{float}
\usepackage{graphicx}
\usepackage{slashed}
\usepackage{multirow}
\usepackage{booktabs}
\usepackage{comment}
\usepackage{ulem}
\usepackage{soul}
\usepackage[fleqn]{mathtools}
\usepackage{dblfnote}
\usepackage{titlesec}

\begin{document}

\def\CB{\textcolor{blue}}
\def\HKD{\textcolor{magenta}} 
\def\RS{\textcolor{red}}
\def\NG{\textcolor{cyan}}
\def\SM{\textcolor{orange}}
\def\AS{\textcolor{green}}

\title{Proposal for a shared transverse LLP detector for FCC-ee and FCC-hh and a forward detector for FCC-hh}

\author{Biplob Bhattacherjee}
\email{biplob@iisc.ac.in}
\affiliation{Centre for High Energy Physics, Indian Institute of Science, Bengaluru 560012, India}

\author{Camellia Bose}
\email{camelliabose@iisc.ac.in}
\affiliation{Centre for High Energy Physics, Indian Institute of Science, Bengaluru 560012, India}

\author{Herbi K. Dreiner}
\email{dreiner@uni-bonn.de}
\affiliation{Bethe Centre for Theoretical Physics and Physikalisches Institut der Universit\"at Bonn, \\Nu\ss allee 12, Bonn 53115, Germany}

\author{Nivedita Ghosh}
\email{nivedita.ghosh@ipmu.jp}
\affiliation{Kavli IPMU (WPI), UTIAS, University of Tokyo, Kashiwa, Chiba, 277-8583, Japan}

\author{\\Shigeki Matsumoto}
\email{shigeki.matsumoto@ipmu.jp}
\affiliation{Kavli IPMU (WPI), UTIAS, University of Tokyo, Kashiwa, Chiba, 277-8583, Japan}

\author{Swagata Mukherjee}
\email{s.mukherjee@cern.ch}
\affiliation{Department of Physics, Indian Institute of Technology, Kanpur 208016, India}

\author{Rhitaja Sengupta}
\email{rsengupt@uni-bonn.de}
\affiliation{Bethe Centre for Theoretical Physics and Physikalisches Institut der Universit\"at Bonn, \\Nu\ss allee 12, Bonn 53115, Germany}

\author{Anand Sharma}
\email{anandsharma2@iisc.ac.in}
\affiliation{Centre for High Energy Physics, Indian Institute of Science, Bengaluru 560012, India}

%\date{}

\begin{abstract}
    As the particle physics community has explored most of the conventional avenues for new physics, the more elusive areas are becoming increasingly appealing. 
    One such potential region, where new physics might be hiding, involves light and weakly interacting long-lived particles (LLPs).
    To probe deeper into this region, where the possibility of highly displaced scenarios weakens the role of general-purpose collider detectors, dedicated LLP detectors are our best option.
    However, their potential can only be fully realized if we optimize their position and dimensions to suit our physics goals. 
    This is possible at the upcoming Future Circular Collider (FCC) facility, where the feasibility and design studies are still ongoing and can accommodate new proposals focused specifically on LLP searches.
    We propose optimized dedicated detectors in both the transverse and forward directions, DELIGHT and FOREHUNT, significantly enhancing the sensitivity to previously uncharted regions of the new physics parameter space.
    Our proposed DELIGHT detector can additionally serve as a shared transverse detector during both the FCC-ee and FCC-hh runs.
    The concept of a shared transverse detector is novel and sustainable, utilizing the same interaction points of the lepton and hadron colliders at the FCC.
    This minimizes costs and boosts the LLP physics case at the FCC.
    %    We propose an optimized dedicated detector in the transverse direction, significantly enhancing the sensitivity to previously unchartered regions of the new physics parameter space.
    %Furthermore, we emphasize the impact of our proposed forward detector, FOREHUNT, along the FCC-hh beam pipe, in probing LLPs from meson decays.
%    \\\vspace{0.5cm}
%    \noindent\rule{0.85\linewidth}{1.5pt}
\end{abstract}

\maketitle

%\tableofcontents

%\clearpage

%\setcounter{page}{1}

%\justifying{}
%\raggedbottom

\section{Introduction}
\vspace*{-0.2cm}

Traditionally, searches for hints of physics beyond the Standard Model (BSM) have primarily focused on TeV-scale particles. 
Nevertheless, these particles have eluded detection so far. 
In light of this, there is increasing interest in exploring light and weakly coupled new physics, particularly in the case of light long-lived BSM particles (LLPs). 
%Most conventional searches for new physics assume that BSM particles decay promptly. 
%However, 
Many BSM models predict the existence of LLPs, including supersymmetric (SUSY) models, dark matter models, models with heavy neutral leptons, and gauge and Higgs portal models.
Various collider searches at ATLAS\,\cite{atlas}, CMS\,\cite{cms}, LHCb\,\cite{lhcb}, and Belle II\,\cite{belleII} target LLPs motivated by different BSM models.
Apart from these, neutrino experiments, such as LSND\,\cite{Foroughi-Abari:2020gju} and MicroBOONE\,\cite{MicroBooNE:2021usw}, and beam dump experiments, like E949\,\cite{BNL-E949:2009dza}, NA62\,\cite{NA62:2020xlg}, PS191\,\cite{Gorbunov:2021ccu}, CHARM\,\cite{CHARM:1985anb}, and KOTO\,\cite{Egana-Ugrinovic:2019wzj}, have contributed significantly to the LLP search program.
The future DUNE\,\cite{Berryman:2019dme} and SHiP\,\cite{SHiP:2015vad} experiments are projected to improve the reach of neutrino and fixed-target experiments, respectively, for light LLPs.
Additionally, dedicated LLP detectors, for example, FASER\,\cite{FASER:2019aik}, are essential to capture highly displaced LLP decays for light boosted LLPs characterized by high mean proper decay lengths.
Several other dedicated LLP detectors are proposed at the LHC, such as MATHUSLA\,\cite{MATHUSLA:2019qpy}, CODEX-b\,\cite{Aielli:2022awh}, and ANUBIS\,\cite{Bauer:2019vqk}, where test runs are envisaged.
%, implying their necessity in the search for LLPs. 

With the ongoing feasibility studies for the Future Circular Collider (FCC)\,\cite{FCC:2018byv,FCC:2018evy,FCC:2018vvp,FCC:2025lpp}, it is timely to investigate the best options for dedicated detectors that could boost its sensitivity to LLPs.
Multiple proposals for dedicated detectors in both the transverse and forward regions of future colliders have been studied\,\cite{Wang:2019xvx,Chrzaszcz:2020emg,Bhattacherjee:2021rml,Schafer:2022shi,Boyarsky:2022epg,Bhattacherjee:2023plj,MammenAbraham:2024gun,Lu:2024fxs,Bhattacherjee:2025dlu}.
This would contribute to strengthening the physics case for the FCC.
In the present study, we highlight the importance of optimized dedicated LLP detectors at the FCC complex, focusing on finding a shared detector concept for both FCC-ee and FCC-hh.

We first aim to determine the region of the parameter space for the dark Higgs model where the HL-LHC loses sensitivity, then evaluate how far the FCC-ee IDEA detector~\cite{FCC:2018byv,FCC:2018evy,Bernardi:2022hny} can extend its reach. 
Finally, we optimize the dedicated detectors at FCC to gain sensitivity to the region beyond the reach of the main collider detector experiments.
While the feasibility of the size and position of these detectors is an important experimental aspect, our goal is to identify the requirements for these detector configurations that are essential to enhance sensitivity compared to general-purpose detectors.

We also propose a novel concept: a dedicated LLP transverse detector whose positioning and size are optimized jointly for FCC-ee and FCC-hh. 
While reusing components from previous colliders has precedent, such as LHC-ALICE's adoption of the LEP-L3 solenoid magnet\,\cite{BUL-NA-2002-143}, reusing an entire dedicated detector is unprecedented.
With proper planning, FCC offers a unique opportunity to pioneer this innovative approach, maximizing resource utilization and promoting sustainability in high energy physics (HEP).

The rest of the paper is organized as: in Sec.\,\ref{sec:role}, we discuss the role of future colliders and introduce the BSM benchmark model we use in this study. In Sec.\,\ref{sec:main_det}, we discuss the reach of the main collider detectors at the HL-LHC and FCC-ee.
In Sec.\,\ref{sec:transverse}, we highlight the importance of the transverse detector, DELIGHT, along with the idea of a shared detector for FCC-ee and FCC-hh. We also study the optimization and minimal configuration of the shared transverse dedicated detector and the possible backgrounds.
In Sec.\,\ref{sec:forward}, we study the role of our FOREHUNT proposal to enhance new physics in the forward direction at the FCC-hh. 
%The optimization and minimal configuration of both the transverse and forward dedicated detectors are studied in Sec.\,\ref{sec:optimise}.
%, and possible timeline and cost estimates are discussed in Sec.\,\ref{sec:timeline}. 
Finally, in Sec.\,\ref{sec:concl}, we conclude. 

\section{Role of Future Colliders in the LLP Sector}
\label{sec:role}
\vspace*{-0.2cm}

The Large Hadron Collider (LHC) has already been approved for upgrades in preparation for the high luminosity run (HL-LHC).
During this period, we will also receive results from the approved DUNE and SHiP experiments. 
If one of the dedicated LLP detector proposals for the HL-LHC, such as MATHUSLA\,\cite{MATHUSLA:2019qpy} or FPF\,\cite{Feng:2022inv}, is approved, it will explore some regions of the LLP parameter space.
After the HL-LHC and other small-scale experiments, we will face two possible scenarios.

%\subsection*{Null observation at LHC}

The first scenario would be if we do not find any hints of new physics 
%If we do not observe any LLP signal 
in the rest of the LHC run, the HL-LHC, or any other approved collider, neutrino, or beam dump experiments.
In such a scenario, the FCC would be our best bet to move ahead.
Owing to a cleaner environment and excellent particle identification, in some cases, the FCC-ee can be sensitive to regions beyond the reach of the HL-LHC.
The FCC-hh, on the other hand, has the huge benefit of larger center-of-mass energy and luminosity, which is expected to outperform the HL-LHC in many aspects.

%\subsection*{Model Identification at FCC from an Observed Signal}

In the second scenario, where we do observe an LLP signal at one or more experiments among the collider experiments, such as HL-LHC, neutrino, or beam dump experiments, the FCC will be crucial in further disentangling the observed signal.
Even if the currently approved experiments observe a signal, FCC-ee/FCC-hh can be sensitive to complementary production or decay modes of the new BSM particle.
They can even play a role in discerning the underlying model and estimating various parameters of the model with a cleaner environment at the FCC-ee or more statistics at the FCC-hh.

\medskip

Having stressed the importance of the FCC initiative, it is also essential to understand the potential of dedicated LLP detectors at the FCC detector complex.
We could gain sensitivity to light and weakly coupled new physics if these dedicated detectors are strategically designed and placed near the various interaction points (IPs) at the FCC.
Unlike the LHC complex, where the dedicated detectors are planned to fit in the available spaces, it is possible to integrate the dedicated detectors at FCC in the design beforehand.
In this regard, it is further appealing to identify whether an optimized LLP detector can be shared by both the future lepton and hadron colliders.

To illustrate the above-discussed ideas and study an optimized and shared detector concept,
we use the dark Higgs model in the our work. The part of the Lagrangian which is relevant for the phenomenology discussed here is as follows:
\begin{equation}
    \mathcal{L} \supset -m_{\phi}^2\phi^2 -\sin\theta\,\frac{m_f}{v} \phi\bar{f}f - \lambda_{h\phi\phi} v\,h\phi\phi,
    \label{eq:Lphi}
\end{equation}
where $m_\phi$ is the mass of the dark Higgs boson, $\theta$ is the mixing angle between the Standard Model (SM) Higgs field and the newly added scalar field, $\lambda$ is the trilinear coupling between the two scalars, $f$ denotes the SM fermions, and $v\approx 246$\,GeV is the vacuum expectation value ({\it vev}) of the Higgs doublet.
The lifetime of the particle $\phi$ is determined by $m_\phi$ and $\sin{\theta}$, and for a major part of the parameter space, $\phi$ is long-lived. 
We examine two potential production modes: $B\to K\phi$ for $m_\phi\in[0.3,4.5]$\,GeV and $h\to\phi\phi$ for $m_\phi\in[6,60]$\,GeV. The branching fraction for the former depends on $\sin{\theta}$, while for the latter, it depends on $\lambda_{h\phi\phi}$. We choose this benchmark model since it is one of the minimal renormalizable portals to light new physics, gives rise to different production and decay modes of the LLP, and the scalar can also act as the mediator to light fermionic dark matter\,\cite{Matsumoto:2018acr}. In addition, it constitutes two recommended
benchmarks, BC4 and BC5, outlined by the Physics Beyond Colliders Collaboration (PBC)\,\cite{Alemany:2019vsk}.

\section{Reach of the main collider detectors at HL-LHC and FCC-ee}
\label{sec:main_det}
% HL-LHC CMS MS
\vspace*{-0.2cm}

Long-lived scalars ($\phi$) produced from the Higgs boson decay are mostly produced in the central region.
Therefore, we use this production mode to illustrate the potential of a transverse LLP detector at FCC.
The LHC general-purpose detectors, like CMS and ATLAS, have already performed searches for LLPs coming from Higgs boson decays, with searches for decays in the muon spectrometer enhancing the sensitivity to much lighter and displaced LLPs\,\cite{CMS:2024bvl,ATLAS:2022gbw,CMS:2021juv,ATLAS:2019jcm}.
We have performed a similar study to estimate the HL-LHC projection for this scenario at the CMS muon spectrometer (CMS MS) in Ref.\,\cite{Bhattacherjee:2021rml}.

\begin{figure}[htb!]
\centering
    \includegraphics[width=0.45\textwidth]{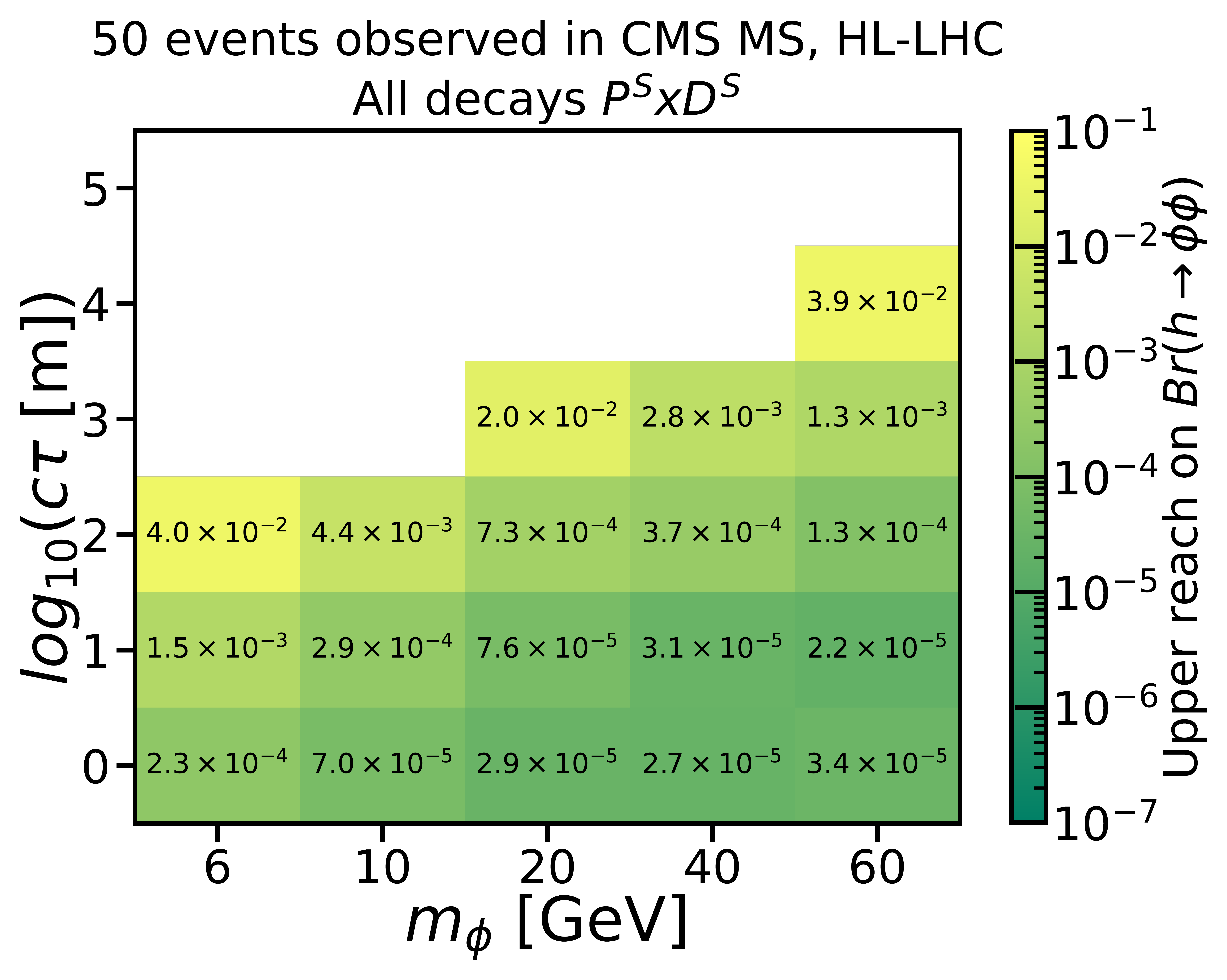}\hspace{1cm}
    \includegraphics[width=0.45\textwidth]{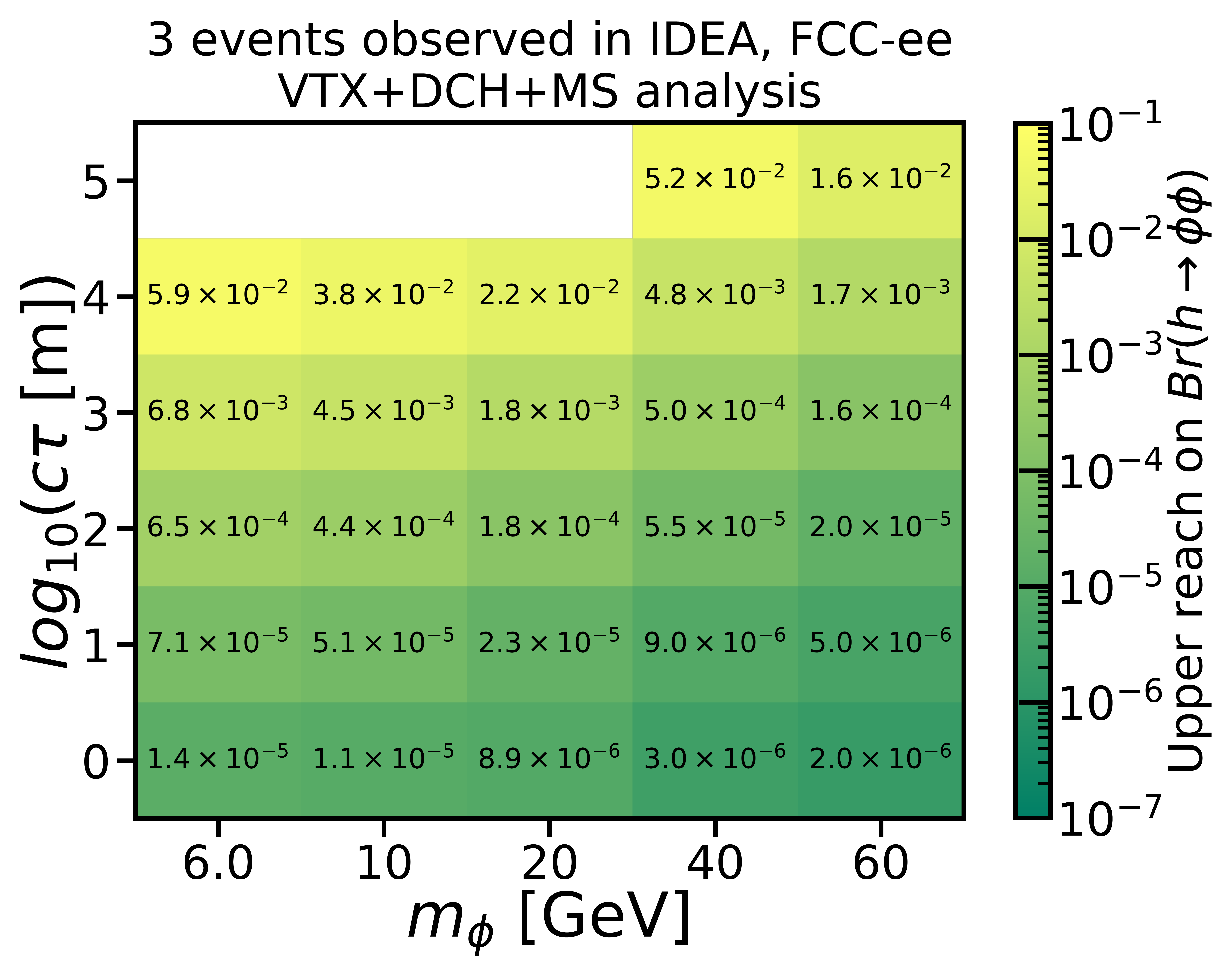}
    \caption{Projected upper reach on the branching fraction, Br($h\rightarrow \phi\phi$), for 50 observed decays of long-lived particles at HL-LHC, using the CMS MS when events are selected by applying the $P^S\times D^S$ cuts of Ref.\,\cite{Bhattacherjee:2021rml} ({\it top}) and for 3 observed decays of long-lived particles at FCC-ee, using the events observed within the vertex detector (VTX), drift chamber (DCH), and the muon system (MS) of FCC-ee, applying the cuts in Table \ref{tab:bkg_ccbb_higgs_cuts} ({\it bottom}).} 
    \label{fig:hl_lhc_fcc_ee}
\end{figure}

At HL-LHC, we consider the Higgs boson production via gluon-gluon fusion (ggF), vector boson fusion (VBF), and the associated production with a vector boson ($Vh$).
The LLPs in the mass range $m_\phi\in[6,60]$\,GeV decay dominantly to $c\bar{c}$ and $b\bar{b}$ final states.
We select events with a prompt object from the Higgs boson production (for example, two forward jets in the VBF mode) and displaced activity from the $\phi$ decay in the CMS MS, where both the prompt and displaced objects satisfy a relatively soft set of cuts, termed as $P^S\times D^S$, defined in Ref.\,\cite{Bhattacherjee:2021rml}. 
The {\it top} panel of Fig.\,\ref{fig:hl_lhc_fcc_ee} shows the resulting upper reach on the branching fraction Br$(h\to\phi\phi)$, for 50 observed events, in the ($m_\phi$, $c\tau$) plane.
The shown sensitivities are obtained by combining the ggF, VBF, and $Vh$ channels for the Higgs boson production at the 14\,TeV HL-LHC (3000 fb$^{-1}$). 

% FCC-ee IDEA
At the FCC-ee run with $\sqrt{s}=240~\text{GeV}$, Higgs bosons are dominantly produced via the $e^-e^+\to Zh$ process.
The Higgs boson subsequently decays to a pair of LLPs.
We follow the analysis strategy discussed in our previous work, Ref.\,\cite{Bhattacherjee:2025dlu}.
Our analysis searches for decays of the LLP within the vertex detector (VTX) and the drift chamber (DCH) for relatively short displacements, and decays in the muon system (MS) for more displaced decays inside IDEA detector (see Refs.\,\cite[Chapter 7]{FCC:2018evy},\cite{FCC:2025lpp}). 
At a center-of-mass energy of $\sqrt{s} = 240$\,GeV, the dominant Standard Model backgrounds that mimic the signal process $e^+e^- \to Zh$, $h \to \phi\phi$ include $e^+e^-\to f\bar{f}$, $WW$, $ZZ$, and $Zh$, where $f$ denotes any SM fermion. We consider both the hadronic and leptonic decays of the vector bosons. 

To analyze the signal and background, we consider several variables characterizing the displaced vertex: the detector element where the vertex lies ($D_{\rm vtx}$), the number of charged particles produced ($N_{\rm ch}$), the total energy of the vertex particles ($E$), transverse displacement from the origin ($d_T$), number of vertices ($N_{vtx}$), invariant mass of the vertex particles ($m_{vtx}$), and the vertex impact parameter ($d_0$).
%Table\,\ref{tab:ZHiggs_bkg_cs} shows the corresponding cross-sections and expected number of events for an integrated luminosity of 10.8\,ab$^{-1}$. 
We simulate $10^8$ events each for $Z\to f\bar{f}$ and $WW$, $2\times10^7$ each for $ZZ$ and $Zh$, and 20,000 events for the signal process.
We select vertices satisfying a certain set of cuts listed in Table\,\ref{tab:bkg_ccbb_higgs_cuts}, and select events having exactly two such vertices for the VTX+DCH analysis and at least one such vertex for the MS analysis. 
The SM background events were reduced to zero with these cuts in both cases.
The {\it bottom} panel of Fig.\,\ref{fig:hl_lhc_fcc_ee} shows the upper reach on the branching fraction Br$(h\to\phi\phi)$ when at least 3 events are observed with the VTX+DCH and MS analyses combined. 

\begin{comment}
\begin{table}[hbt!]
\centering
\begin{tabular}{|c|c|c|}
\hline
SM Backgrounds & Cross-section (pb) & Expected events\\ \hline\hline
$e^+e^-\to f\bar{f}$ & 21.43 & $2.3\times 10^8$\\
$e^+e^-\to W^+W^-$ & 16.84 & $1.8\times 10^8$\\
$e^+e^-\to ZZ$ & 1.4 & $1.5\times 10^7$\\
$e^+e^-\to Zh$ & 0.237 & $2.5\times 10^6$\\ \hline\hline
\end{tabular}
\caption{SM backgrounds with their cross-sections and expected number of events at $\sqrt{s}$ = 240 GeV, $\mathcal{L} = 10.8$\,ab$^{-1}$\AS{\cite{FCC:2018evy}}.}
\label{tab:ZHiggs_bkg_cs}
\end{table}
\end{comment}

\begin{table}[hbt!]
    \centering
    \begin{tabular}{|c|c|}
    \hline
        \multicolumn{2}{|c|}{Cuts on background in the $c\bar{c}$ and $b\bar{b}$ final states} \\ \hline\hline
        VTX+DCH analysis & MS analysis \\ \hline
        $D_{\rm vtx} \in$ VTX or DCH &$D_{\rm vtx} \in$ MS \\
        $N_{\rm ch} \ge 3$ & $N_{\rm ch} \ge 3$ \\
        $E > 5$\,GeV & $E > 5$\,GeV\\
        $d_T > 5$\,mm & $d_T > 100$\,mm \\ 
        $d_0 > 5$\,mm & $d_0 > 5$\,mm\\
        $m_{vtx} > 2$\, GeV & $m_{vtx} > 2$\, GeV \\
        $N_{\rm vtx} = 2$  & $N_{\rm vtx} \geq 1$\\
        
        \hline\hline
    \end{tabular}
    \caption{Selection cuts applied on the signal $e^+e^-\to Zh, Z\to incl., h\to \phi\phi$ and the SM background when analysis is done in the $c\bar{c}$ and $b\bar{b}$ final states.}
    \label{tab:bkg_ccbb_higgs_cuts}
\end{table}
%\CB{Although the search is background-free at FCC-ee, we maintain the requirement of at least 10 observed events to keep the results statistically reliable.}

We observe that, based on this preliminary study, the FCC-ee analysis improves the sensitivity compared to the HL-LHC CMS MS analysis.
The best sensitivity on the reach of Br($h\to\phi\phi$) with both analyses is achieved for a LLP of mass 60\,GeV with $c\tau=1$\,m, with the upper reach on the branching fraction improving from $3.4\times10^{-5}$ at the HL-LHC to $2.0\times10^{-6}$ at the FCC-ee.
Moreover, FCC-ee can probe lighter and more displaced regions of the ($m_\phi$, $c\tau$) plane previously beyond the CMS MS analysis at the HL-LHC.
Our analysis has some sensitivity to decay lengths of $10^4\,\text{m}$ for all masses of the LLP, and even up to $10^5\,\text{m}$ for $m_{\phi} = 50$ and $60\,\text{GeV}$ with an upper reach of Br$(h\to\phi\phi)\sim 6\times 10^{-2}$.
However, we still lack sensitivity in the further light and displaced region of the parameter space, particularly for $m_\phi\lesssim 20$\,GeV having $c\tau\sim 10^5$\,m. 
%and $m_\phi\lesssim 40$\,GeV having $c\tau\sim 10^5$\,m.
Subsequently, we focus on dedicated LLP detectors for probing this region.

\section{DELIGHT: a shared transverse detector concept}
\label{sec:transverse}
\vspace*{-0.2cm}
%DELIGHT
In Ref.\,\cite{Bhattacherjee:2021rml}, we proposed the DELIGHT-B ({\bf De}tector for {\bf l}ong-l{\bf i}ved particles at hi{\bf gh} energy of 100 {\bf T}eV) transverse LLP detector for the FCC-hh experiment, placed 25\,m from the FCC-hh IP and having a decay volume of $100\times 100\times 100$\,m$^3$.
We found that the performance of such a large detector placed relatively close to the IP significantly enhanced the sensitivity.
The improvement partially results from the huge gain in the Higgs production cross-section at the FCC-hh. 
Additionally, aligning the placement of the detector with the direction of the $\phi$ production, {\it i.e.,} placing it at the central pseudorapidity ($\eta$) region, enhances its geometric acceptance.

\begin{figure}[hbt!]
    \centering    
    \includegraphics[width=0.7\textwidth]{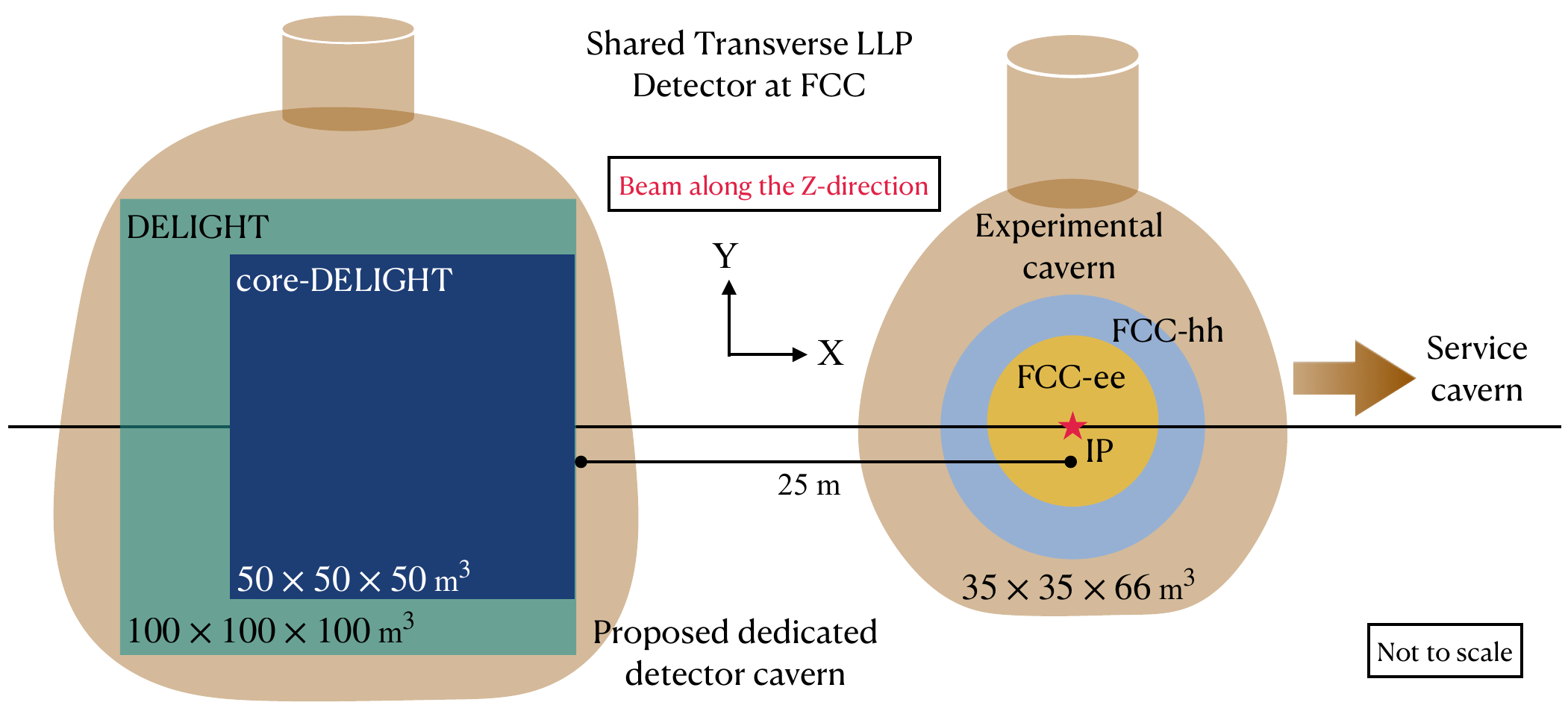}
    \caption{Schematic illustration of the proposed shared transverse LLP detector concept (DELIGHT) at the FCC complex. The detector is placed at 25\,m away from the interaction point in the transverse ($X$) direction, with the beam axis along the $Z$ direction. Two configurations are indicated -- the full DELIGHT detector with a decay volume of $100 \times 100 \times 100~\mathrm{m}^3$, and the minimal configuration (core-DELIGHT) with a volume of $50 \times 50 \times 50~\mathrm{m}^3$, obtained from the optimization study. The existing experimental cavern and a proposed dedicated cavern for the LLP detector are also shown for context. The same detector concept is proposed to be shared between FCC-ee and FCC-hh runs.}
    \label{fig:FCC_LLP_det}
\end{figure}

Since the FCC-ee collider is proposed to precede the FCC-hh collider, we estimate the reach of the DELIGHT-B detector if it is already deployed during the FCC-ee run in Ref.\,\cite{Bhattacherjee:2025dlu}. Henceforth, we drop the categorization of Ref.\,\cite{Bhattacherjee:2021rml}, and refer to the DELIGHT-B configuration simply as DELIGHT.
It performs best among a large class of considered designs and dimensions of dedicated detectors at the FCC-ee.
This gives rise to the idea of a shared detector for both the FCC-ee and the FCC-hh experiments.
According to the planned design of the FCC complex, the FCC-ee and FCC-hh are expected to share the same experimental cavern with the IPs at the same location\,\cite{FCC:2025lpp,FCC:2025uan}.
%of the FCC-ee and the FCC-hh are expected to be $\sim 10$\,m apart\,\cite{FCC:2018evy}. 
Originally DELIGHT detector was proposed to be placed at an elevated position in the FCC-hh tunnel, {\it i.e.,} 25\,m above the FCC-hh IP.
If constructing a huge detector within the experimental cavern proves to be unfeasible in reality, we propose that the detector be placed in a separate cavern, 25\,m away in the $X$-direction from the FCC-ee and FCC-hh IPs, and symmetrically extending in the $Y$ and $Z$-directions. 
Since the collision events are symmetric in the azimuthal plane, changing the detector's $X$-$Y$ coordinates preserves efficiency.
%Thus, this shared LLP detector will be 35\,m away from the FCC-hh IP.
%Moreover, moving it closer to the FCC-ee IP is preferred since the gain in the production cross-section of Higgs bosons at FCC-hh would compensate for its larger distance from DELIGHT. 
We show an illustration of the proposed shared DELIGHT detector in Fig.\,\ref{fig:FCC_LLP_det}.

%The proposed detectors for dedicated LLP searches are, namely, the transversely positioned detector DELIGHT and the forwardly positioned detector FOREHUNT. Fig.\,\ref{fig:FCC_LLP_det} illustrates the placement and configurations of these detectors. We explore the design and sensitivity of each detector in the following sections.
  
To compute the sensitivity of DELIGHT placed at a modified position with respect to the FCC-hh, we generate Higgs boson production via the ggF channel at a center-of-mass energy of 100\,TeV.
This results in an enhancement in the production cross-section by $\sim 3000$ times compared to FCC-ee and $\sim 16$ times compared to HL-LHC. 
FCC-hh is planned to collect an integrated luminosity of $30\,\text{ab}^{-1}$\,\cite{FCC:2018vvp,FCC:2025lpp}, which is thrice the luminosity of FCC-ee during the $Zh$ run and $10$ times larger than the luminosity at the end of HL-LHC.
With the new configuration of DELIGHT, we calculate the upper limits on $BR(h\to \phi\phi)$ for $m_\phi\in[6,60]$\,GeV when at least 3 LLP decays are observed within the DELIGHT detector with the energy of the LLP, $E > 5\,\text{GeV}$. 
Before discussing the results, we first examine the optimization of the detector’s position and dimensions in order to determine the minimal configuration of the shared transverse detector.

%%% optimization

\subsection{Optimization and the core-DELIGHT}
\vspace*{-0.2cm}

We seek to find a balance between the detector's performance and the practicality of its position and dimensions.
It is expected that a detector close to the IP with a large cross-sectional area towards the IP and a large decay volume would have the best performance; however, it might not be practical to construct and deploy.
This requires optimization to determine how much compromise on the performance comes from a smaller and farther dedicated detector, while still providing a gain compared to general-purpose detectors.  
For box-shaped detectors, the performance of the detector depends on three parameters, namely, the length of the detector $(L)$, the surface area of the detector $(A)$, and the distance from the FCC-ee IP $(D)$. 
The detectors are moved away from the FCC-ee IP in steps of $2\,\text{m}$, starting from a distance of 20\,m.
We consider 4100 different configurations by changing these three parameters and calculate the efficiency of the detectors for the following nine benchmarks in the $(m_{\phi},~c\tau)$ parameter space:
\begin{itemize}
    \itemsep-0.3em
    \item (6\,GeV, 10\,m), (6\,GeV, 10$^3$\,m), (6\,GeV, 10$^5$\,m)
    \item (20\,GeV, 10\,m), (20\,GeV, 10$^3$\,m), (20\,GeV, 10$^5$\,m)
    \item (60\,GeV, 100\,m), (60\,GeV, 10$^3$\,m), (60\,GeV, 10$^5$\,m)
\end{itemize}
These benchmarks are chosen so that the efficiency of observing them in DELIGHT is at least twice that in the FCC-ee IDEA detector. 
For each of these different detector configurations, we calculate the ratio of the detector efficiency, $\epsilon_{\rm det}$, to the efficiency of the FCC-ee IDEA detector, $\epsilon_{\rm IDEA}$, for the nine benchmarks. The efficiency, $\epsilon_{\rm IDEA}$, denotes the geometrical acceptance of the entire IDEA detector, representing the probability that at least one of the LLP decays to charged final state within the detector’s volume. 
%Modern machine learning techniques may be able to improve the efficiency of searches at the main detector and hence further lower the gain achieved with a far detector.
We compute the average, which gives us the overall gain obtained from the dedicated detector over FCC-ee.
We plot the gain, $G_T^{\rm det}$ defined below, as a function of $D$, for three fixed values of $A$ and $L$ each, in Fig.\,\ref{fig:transverse_opt}.
\begin{equation}
    G_T^{\rm det} = \frac{1}{9}\sum_{\rm benchmarks} \frac{\epsilon_{\rm det}}{\epsilon_{\rm IDEA}}
\end{equation}

\begin{figure*}[t!]
    \centering
    \includegraphics[width=0.32\textwidth]{A10000.pdf}
    \includegraphics[width=0.32\textwidth]{A2500.pdf}
    \includegraphics[width=0.32\textwidth]{A400.pdf}
    \caption{The gain in efficiency of the dedicated LLP transverse detectors over the main detector as a function of the distance $(D)$ from the IDEA IP for three different lengths $(L)$ of the detectors. \textit{From top to bottom:} the surface area of the transverse detectors are respectively fixed at $100\times 100~\text{m}^2$, $50\times 50~\text{m}^2$ and $20\times 20~\text{m}^2$. The {\it dashed line} represents the gain of the DELIGHT detector. The {\it red dots} mark the position $D$ at which the dedicated LLP detectors perform exactly as the IDEA detector for all the benchmarks. If the detectors are placed beyond these positions, they lose sensitivity (gain $<$ 1) for a few individual benchmarks.}
    \label{fig:transverse_opt}
\end{figure*}

We observe that the gain decreases rapidly with increasing $D$. 
If a gigantic detector with $10^6\,\text{m}^3$ volume can be built, even at 100\,m from the IP, it shall be $\sim 3$ times more efficient than the FCC-ee IDEA detector in capturing LLPs.
However, we must remember that this is only an average gain over nine benchmark points.
Some benchmarks with high masses and low decay lengths, for example, the (20\,GeV, 10\,m) and (60\,GeV, 100\,m) LLP benchmarks, may escape detection, while benchmarks with higher decay lengths may compensate for the loss. 
We calculate the positions of the dedicated detectors where all the benchmarks can be observed with at least the same efficiency as the FCC-ee IDEA detector. 
Such positions are marked with {\it red dots} in Fig.\,\ref{fig:transverse_opt}. 
For the $A=100\times 100~\text{m}^2$ configuration, the positions are 42\,m, 40\,m, and 34\,m, for detectors with $L=100,~50$, and 20\,m respectively. 
For a smaller configuration, where $A=50\times 50~\text{m}^2$, we must place the detector closer, at 26\,m when $L=100$ and 50\,m. 
The detectors with $A=20\times 20~\text{m}^2$ are too small to observe all benchmarks with equal sensitivity as the IDEA detector.
This demonstrates that beyond a certain point, compromising on the detector dimensions and position would not yield any benefit over the IDEA detector at FCC-ee.
We identify the minimal transverse detector design, having a decay volume of $50\times50\times50$\,m$^3$ and placed 26\,m away from the FCC-ee IP, as the core-DELIGHT configuration (see Fig.\,\ref{fig:FCC_LLP_det}). 

%%%% results

\begin{figure}[htb!]
    \centering
    \includegraphics[width=0.49\textwidth]{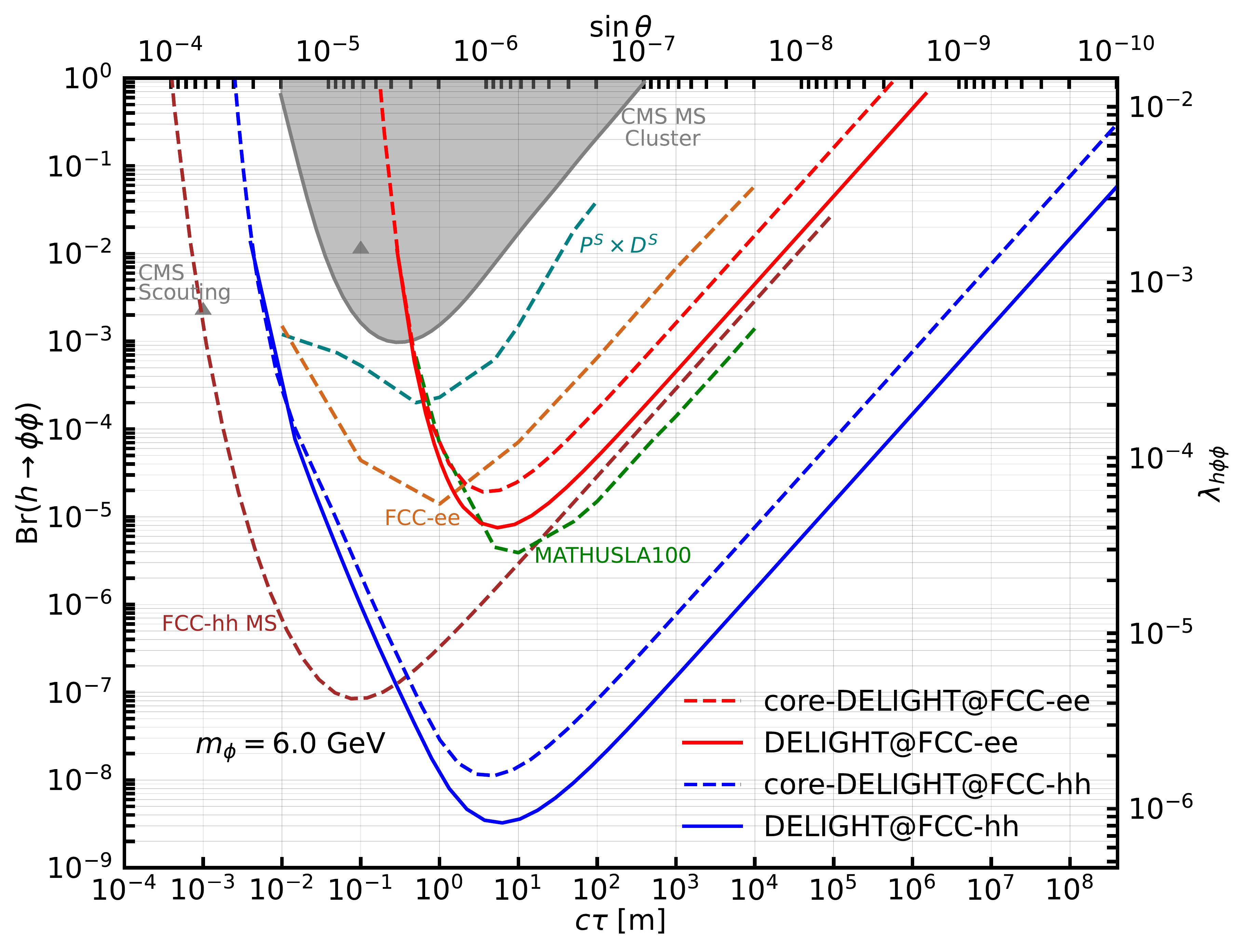}
    \includegraphics[width=0.49\textwidth]{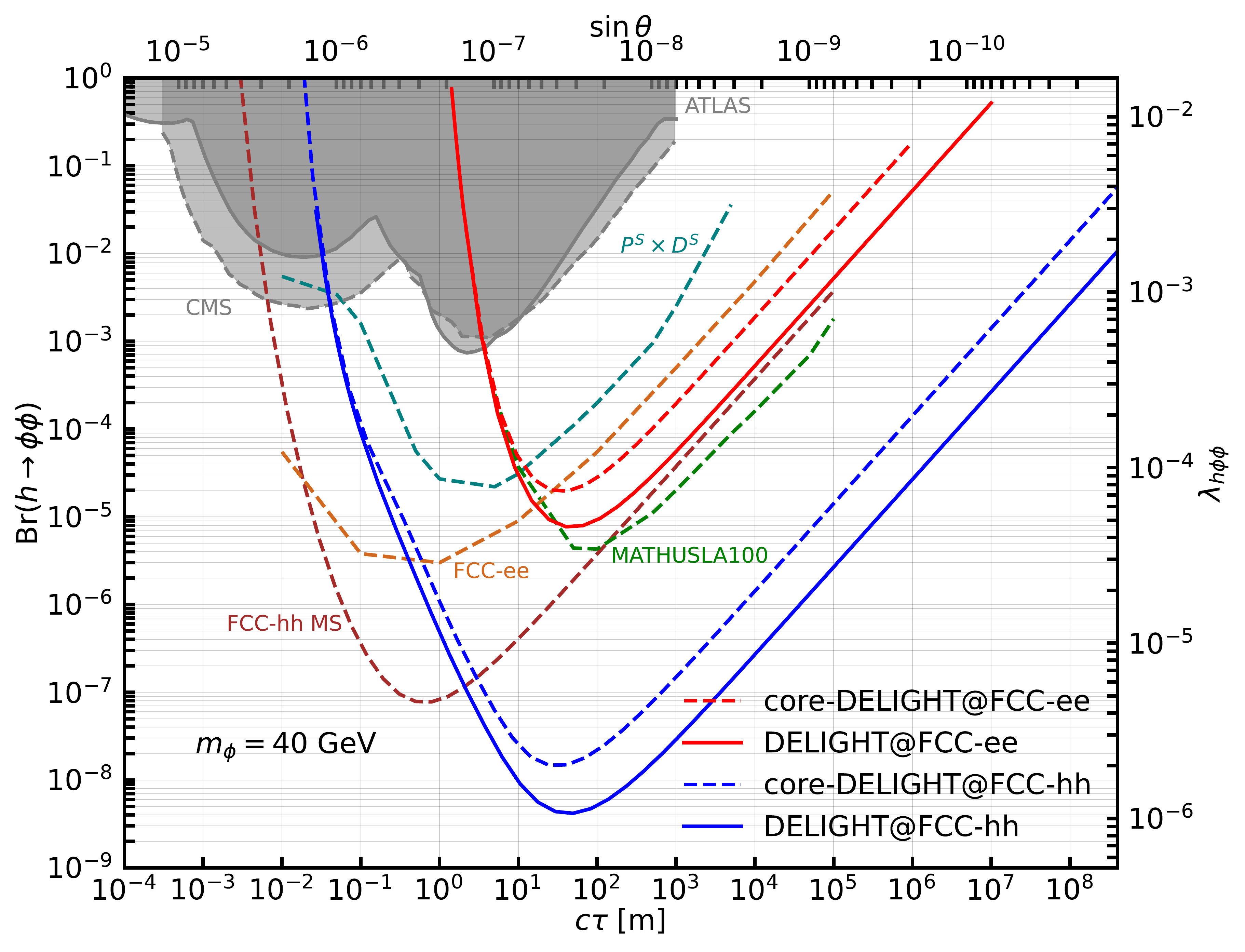}
    \caption{Projected upper limits on the branching fraction Br($h \rightarrow \phi\phi$), assuming 3 observed long-lived particle decays within the DELIGHT (and core-DELIGHT) detector. Results are shown for LLP production at FCC-ee ({\it red}) and FCC-hh ({\it blue}), for a 6\,GeV ({\it left}) and a 40\,GeV ({\it right}) LLP. For comparison, we also show the current ATLAS and CMS bounds ({\it grey}), as well as the projected sensitivities of the CMS MS at HL-LHC ($P^S \times D^S$), MATHUSLA at HL-LHC (both from Ref.\,\cite{Bhattacherjee:2021rml}), the FCC-ee IDEA detector, and the FCC-hh MS. We also show the corresponding underlying model parameters, $\sin \theta$ and $\lambda_{h\phi\phi}$, in the {\it top} and {\it right} axes of the plots respectively.}
    \label{fig:delight_fcc_ee_hh}
\end{figure}

The results for the shared core-DELIGHT and DELIGHT detectors at the FCC-ee and FCC-hh are shown for $m_\phi=6$\,GeV ({\it top}) and $m_\phi=40$\,GeV ({\it bottom}) in Fig.\,\ref{fig:delight_fcc_ee_hh}, with both the detectors placed at 25\,m for both the runs.
We also show the present limits from the ATLAS and CMS searches\,\cite{summary}.
The projected sensitivities from the CMS MS ($P^S\times D^S$ with 50 observed events) and MATHUSLA (with 3 events) shown in Fig.\,\ref{fig:delight_fcc_ee_hh} are taken from Ref.\,\cite{Bhattacherjee:2021rml}, while the projected sensitivity for the FCC-ee IDEA detector is estimated using the analysis described in Section\,\ref{sec:main_det} assuming 3 observed events.
We also estimate the projected sensitivity of the muon spectrometer search at the FCC-hh. In order for an LLP event to be triggered in the muon spectrometer, we require that at least one of the $\phi$ particles has a transverse momentum of at least 200\,GeV\,\cite{PhysRevD.108.055040}, which effectively suppresses SM backgrounds. To obtain such a high-$p_T$ spectrum for the LLPs, we generate events with Higgs bosons along with a hard jet using \textsc{Pythia8} at $\sqrt{s}=100$\,TeV with a \texttt{HardQCD} process and a minimum transverse momentum transfer of $\hat{p}_T\ge$500\,GeV. The generated Higgs bosons are then decayed to a pair of $\phi$, and the resulting LLP kinematic distributions are used to derive upper limits on BR($h\rightarrow\phi\phi$) at 30\,ab$^{-1}$ of integrated luminosity.
The branching ratio $\mathrm{BR}(h \rightarrow \phi\phi)$ is directly controlled by the coupling $\lambda_{h\phi\phi}$, and can be written as\,\cite{Feng:2017vli}
\begin{equation}
\mathrm{BR}(h\to\phi\phi)
\approx \frac{\lambda_{h\phi\phi}^2\,v^2}{8\pi\,m_h\,\Gamma_\text{SM}}
\sqrt{1 - \frac{4m_\phi^2}{m_h^2}}\,,
\end{equation}
where we have used standard properties of the Higgs boson (e.g.\ $\Gamma_\text{SM}\simeq 4.07~\mathrm{MeV}$). For a fixed LLP mass, this provides a one-to-one mapping between $\mathrm{BR}(h \rightarrow \phi\phi)$ and $\lambda_{h\phi\phi}$.
Similarly, the LLP lifetime is governed by the mixing angle $\sin\theta$, with $c\tau \propto 1/\sin^2\theta$, up to a mass-dependent factor determined by the decay widths. The relation between $c\tau$, $\sin\theta$, and $m_\phi$ has been taken from Ref.\,\cite[Fig.\,1]{Bhattacherjee:2025dlu}.
We use these relations to directly translate the $c\tau$-dependent upper reach on Br($h\to\phi\phi$) for the fixed masses of $\phi$ into the underlying theory parameters, $\sin\theta$ and $\lambda_{h\phi\phi}$, also shown in Fig.\,\ref{fig:delight_fcc_ee_hh}.
%The results for FCC-ee and FCC-hh with the shared DELIGHT detector are shown in the {\it top} and {\it bottom} panels of Fig.\,\ref{fig:delight_fcc_ee_hh} respectively.

On comparing the various sensitivity curves in Fig.\,\ref{fig:delight_fcc_ee_hh}, we observe that DELIGHT is sensitive to the region with relatively high decay lengths and performs slightly better for low masses. 
Additionally, among the benchmarks that were already observed at the IDEA detector, we gain in sensitivity by a factor of around 7 for ($m_{\phi},~c\tau$) =($40\,\text{GeV},~1000\,\text{m}$) to a factor of 14 for ($m_{\phi},~c\tau$) =($6\,\text{GeV},~1000\,\text{m}$).
For the $40$\,GeV LLP having $c\tau\lesssim 10\,\text{m}$, the FCC-ee analysis with the IDEA detector provides better sensitivity than the dedicated detectors. 
Note that we show here the projected sensitivity of the MATHUSLA100 configuration. However, in their input submitted to the EPPSU, they propose a MATHUSLA40 configuration with a decay volume of $40\times40\times11$\,\cite{MATHUSLA:2025eth}, which can probe a branching fraction up to $10^{-4}$. If this MATHUSLA configuration is approved, both the FCC-ee IDEA and DELIGHT detectors will extend the reach on Br($h\to\phi\phi$) by an order of magnitude compared to MATHUSLA. 
During the FCC-hh run, the shared detector DELIGHT will be able to probe BR$(h\to \phi\phi)$ up to $10^{-8}$.\\

\begin{figure}[hbt!]
\centering
\includegraphics[width=0.5\textwidth]{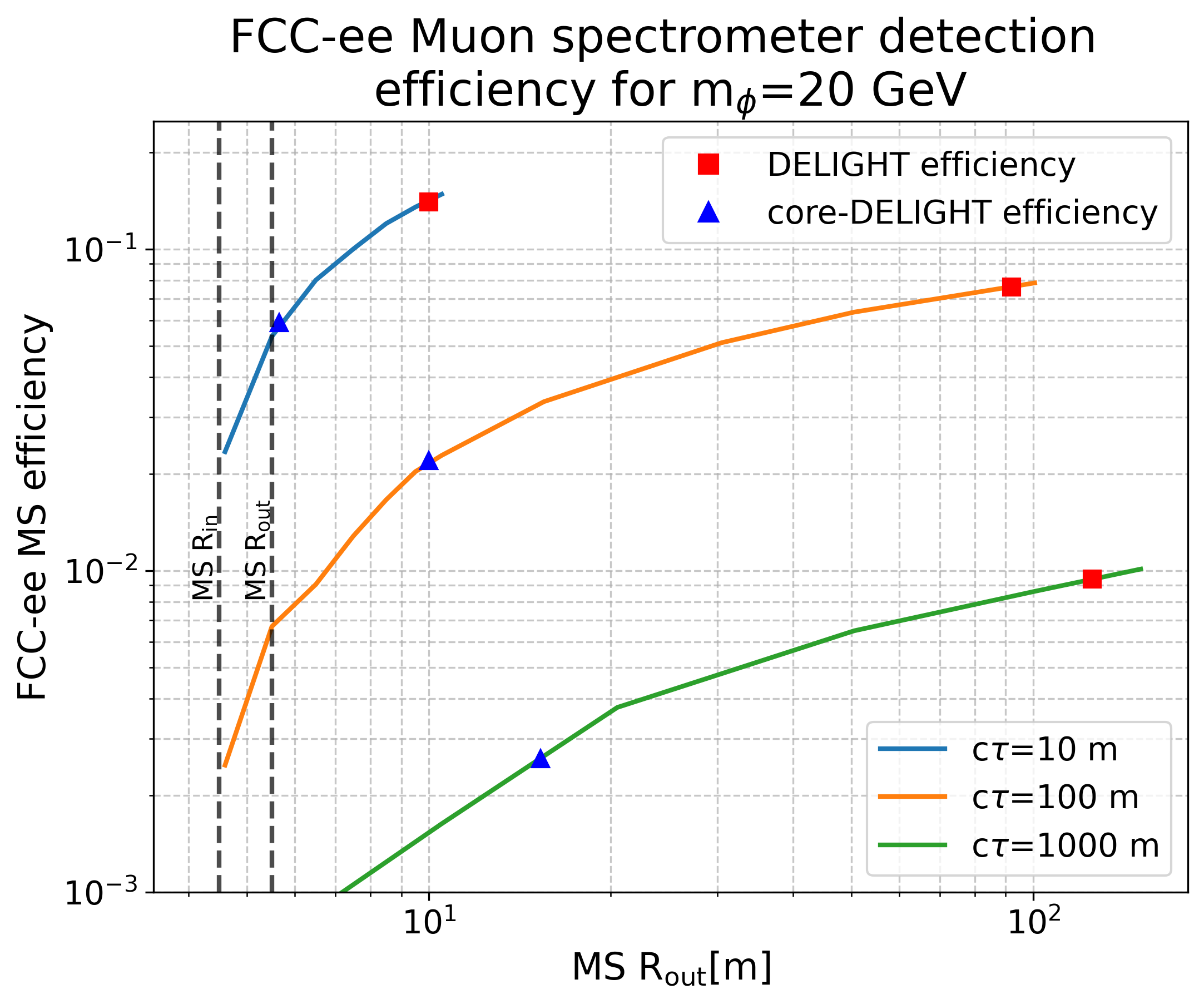}% plot
\caption{Detection efficiency of an extended FCC-ee muon spectrometer for the benchmark process \(h \rightarrow \phi\phi\) as a function of the outer radius of the muon spectrometer, \(R_{\text{out}}\), with the inner radius fixed at \(R_{\text{in}} = 4.5\,\mathrm{m}\) (indicated by the vertical dotted line). Results are shown for a fixed LLP mass of \(m_\phi = 20\,\mathrm{GeV}\) and for three representative lifetimes, \(c\tau = 10\,\mathrm{m}, 100\,\mathrm{m}\), and \(1000\,\mathrm{m}\). For comparison, the efficiencies of core-DELIGHT and DELIGHT are marked on the curve to show exactly where the extended MS reaches equivalent performance for a given mass and $c\tau$.}
\label{extended_MS} 
\end{figure}

To assess whether extending the central detector -- specifically the muon spectrometer (MS) -- could provide a comparable physics gain to a dedicated LLP detector, we performed a dedicated study for the FCC-ee case. We have evaluated the LLP detection efficiency of an extended muon spectrometer for the benchmark process $h\rightarrow\phi\phi$, using decays within the muon spectrometer.
In this study, the inner radius of the muon spectrometer was fixed to its nominal value of R$_{\text{in}}$=4.5\,m,\,\cite[Chapter 7]{FCC:2018evy},\cite{FCC:2025lpp} while the outer radius R$_{\text{out}}$ was varied from its baseline value of 5.5\,m to larger radii. The MS efficiency was evaluated as a function of R$_{\text{out}}$ for a fixed LLP mass of m$_{\phi}$=20\,GeV and for three representative lifetimes, $c\tau$=10\,m, 100\,m, and 1000\,m. The results are shown in Fig.\,\ref{extended_MS}. For comparison, the corresponding efficiencies of core-DELIGHT and DELIGHT for the same mass and lifetimes are also shown in Fig.\,\ref{extended_MS} below.

%\RS{Two key features emerge from this comparison. First, increasing the outer radius of the muon spectrometer leads to only a very slow improvement in efficiency for a fixed LLP lifetime. Second, as the LLP lifetime increases, achieving efficiencies comparable to those of DELIGHT or core-DELIGHT would require extending the muon spectrometer to impractically large radii. 
We observe that while modest MS extensions may provide some sensitivity gain for LLPs with short lifetimes, achieving efficient coverage of long-lived scenarios would require substantially larger MS extensions in the central detectors. Consequently, extending the muon spectrometer alone cannot match the physics reach of a dedicated LLP detector. A purpose-built detector, such as DELIGHT or core-DELIGHT, is therefore necessary to achieve significant sensitivity to LLPs with large decay lengths.

The proposal entails the excavation of an additional cavern to accommodate the detector, which would represent a major component of the construction budget. To give a perspective on the feasibility of constructing such a large cavern close to the experimental cavern, we take the example of the DUNE experiment. The far detector of the DUNE experiment, currently under construction, has dimensions of approximately 150\,m$\times$20\,m$\times$28\,m and is located at a depth of about 1.5\,km\,\cite{heise2021stanford}. The core-DELIGHT cavern has to be larger by roughly a factor of 1.5. The DUNE caverns at the far site are also quite close to each other with rock pillars as structural support. Another major experiment currently under construction is Hyper-Kamiokande, which will feature a cylindrical tank 71\,m in height and 68\,m in diameter \cite{HyperKDetector}. Located at a depth of 600\,m, it is set to become the world’s largest underground water Cherenkov detector. The total volume of this tank is approximately double that of the core-DELIGHT geometry. Given that such large-scale underground constructions are already being realized today, we consider it reasonable to remain cautiously optimistic that comparable or larger engineering challenges could be addressed for a far-future experiment such as FCC.

We also recognize that such extensive civil engineering work could have significant environmental implications. Again we take precedence from the excavation of the large underground caverns for the DUNE experiment at Fermilab, where the 2015 environmental assessment concluded that the project would have no significant impact\,\cite{LBNF_EnvAssessment}. A similarly detailed study would be essential to evaluate the feasibility and the potential environmental consequences of constructing a large cavern adjacent to the FCC experimental cavern. From the physics perspective, however, we argue that this location would be optimal for hosting a shared detector for both FCC-ee and FCC-hh.

While reusing the same cavern for FCC-ee and FCC-hh operation, the detector requirements are indeed expected to differ, primarily due to the different background environments and different collision rates ($\sim$\,20\,MHz for FCC-hh and $\sim$\,100\,KHz for FCC-ee)\,\cite{FCC:2018vvp,FCC:2018evy,FCC:2025uan}.
In principle, the core detector technologies, such as the charged particle detection using Resistive Plate Chambers (RPCs), the triggering algorithms to reject backgrounds, and the choice of detector gas, can remain unchanged between the two running modes.
%The DELIGHT detector could use layers of Resistive Plate Chambers (RPCs) for detecting the charged particles from the LLP decay. 
RPCs typically operate with a gas mixture containing fluorinated gases, which have high global warming potentials. However, researchers are exploring eco-friendly alternatives~\cite{Rigoletti:2023aop}. To minimize environmental impact, DELIGHT should prioritize using these sustainable alternatives, even if this incurs a slight increase in cost.

The main difference in the FCC-ee and FCC-hh running mode will arise in the shielding requirements and triggering thresholds. For the FCC-hh operation, DELIGHT is expected to require increased shielding due to the significantly higher background levels at FCC-hh. We discuss the potential backgrounds in the next section, however, we keep a more detailed background study for a future work.
%would necessitate substantially thicker shielding, of the order of at least 5–6\,m, to adequately suppress additional backgrounds.

\subsection{Backgrounds}
\vspace*{-0.2cm}

%Geant4 simulations have demonstrated that 
%The DELIGHT detector will primarily encounter backgrounds from high-energy muons and neutrinos, in addition to low-energy neutrons, electrons, and photons. The main detectors of the FCC will be instrumental in mitigating other types of background interference. Addressing these backgrounds will necessitate the implementation of several mitigation techniques.

The DELIGHT detector will primarily encounter backgrounds from high-energy muons and neutrinos originating from the FCC IP. In addition, there will be backgrounds from low-energy neutrons, electrons, and photons, from secondary interactions of muons and neutrinos beyond the FCC main detectors.
Addressing these backgrounds necessitates the implementation of several mitigation techniques.
For handling the backgrounds at DELIGHT, we closely follow the approach outlined in the CODEX-b experiment\,\cite{gligorov2018searching}.

Firstly, proximity of DELIGHT to the main FCC detectors make its integration with the trigger system of the FCC experiments feasible. This will enable us to leverage timing constraints to effectively reduce low-energy background events. 
Secondly, we suggest employing several meters of lead and concrete shielding before the DELIGHT detector. This shielding will attenuate or stop electrons, neutrons, and some muons, and cause them to fail the timing constraints. However, a substantial number of high-energy muons will still penetrate this shielding. These muons will be vetoed using tracking information from the MS of the FCC detectors. Additionally, one can add scintillating layers before the decay volume of the DELIGHT detector, which can veto any charged particles coming from the IP.

Finally, neutrinos can be a potential source of background at DELIGHT, which interact within the detector and hence, cannot be vetoed using the scintillating layers. The dominant neutrino background will come from $Z$ boson decays during the $Z$-pole run of the FCC-ee. The FCC-ee is projected to produce approximately $5\times10^{12}$ $Z$ bosons\,\cite{abada2019fcc}, which decay into neutrinos with a 20\% branching ratio. The estimated number of neutrinos impinging on DELIGHT will therefore be,
\begin{align*}
 N_{\nu}&=N_Z\times BR(Z\rightarrow\nu \Bar{\nu})\times\epsilon_{\text{DELIGHT}}, 
\end{align*}
where $\epsilon_{\text{DELIGHT}}$ represents the geometrical efficiency of DELIGHT. We estimate this efficiency to be 27\%, thereby,
\begin{align*}
 N_{\nu}&= 5\times 10^{12}\times0.2\times 0.27\\
 N_{\nu}& \sim 0.3 \times 10^{12}.
\end{align*}
Assuming DELIGHT is filled with gas at Normal Temperature and Pressure (NTP), the incident neutrinos can interact with the gas with a cross-section of approximately 0.01\,pb $\times E_{\nu}$ GeV\,\cite{formaggio2012ev}, and produce charged particles which can mimic an LLP signal. The total number of such interaction events can then be calculated as,
\begin{align*}
      N_{\text{events}}&=N_{\nu}\times n_{\text{gas}}\times \sigma\times L_\text{{DELIGHT}},
\end{align*}
where $L_{\text{DELIGHT}}$ is the length of the DELIGHT detector, which is 100\,m;
$n_{\text{gas}}$ is the number density of the gas within DELIGHT, approximately $3\times10^{25}$ particles/m$^3$ at NTP; and $\sigma$ is the neutrino-nucleon cross-section, which for a 10\,GeV neutrino is 0.1\,pb.
Substituting these values, we get,
\begin{align*}
    N_{\text{events}}&= 0.3 \times 10^{12} \times 3\times10^{25}\text{particles/m}^3 \times 0.1\text{pb}\times 100\text{m}\\
    N_{\text{events}}& \sim 0.01.
\end{align*}
Since $N_{\text{events}}\ll 1$, the neutrino background can be effectively ignored for DELIGHT at FCC-ee.
For the $Zh$ run, where we expect most of the LLP signal coming from Higgs boson decays, the neutrino background is estimated to be significantly lower than that of the $Z$-pole run. Given an integrated luminosity of $5\,\text{ab}^{-1}$ and a cross-section of approximately $200\,\text{fb}$ for $Zh$ production \cite{abada2019fcc}, the total number of events is expected to be $\sim 10^6$. This represents a suppression factor of $10^6$ relative to the total event count of the $Z$-pole run. Consequently, even if every $Zh$ event were to produce a neutrino, the total number of neutrino interactions within the DELIGHT detector would remain negligibly small (well below unity). Hence, the neutrino background in the $Zh$ production peak run at the FCC-ee is also computed to be negligible.

Estimating the neutrino background at the FCC-hh is more involved with multiple sources of neutrino production. These neutrinos arise from the decay of various mesons, $W$ and $Z$ bosons, and top quarks, and their origin stems from both soft and hard-QCD processes. Given the high-luminosity environment of the FCC-hh, pileup will also be an important factor in these calculations. We estimate that the DELIGHT detector will observe $\mathcal{O}(1000)$ neutrino events over the full $30\text{ ab}^{-1}$ integrated luminosity. These events represent a potential background that may mimic the LLP signal. Given that the primary contribution arises from \texttt{softQCD} processes, the majority of these neutrinos have low energies. Consequently, developing dedicated discrimination techniques will be essential to identify the signal over this background. While neutrinos may also interact with the surrounding structural materials, potentially contributing to the background, the DELIGHT detector could be equipped with an active veto on its front face. This system enables efficient rejection of such interactions occurring in the surrounding material. However, further detailed studies are necessary to quantify the rates of these secondary interactions and their potential impact on signal sensitivity.

%\sout{It is anticipated that the detector will register a significant number of neutrino events, some reaching energies of several TeV. As the neutrino-nucleon interaction cross-section scales with energy, a detailed calculation of event rates per energy bin is necessary. However, such an extensive study is beyond the scope of this paper, and we keep it for a future study.
%As mentioned earlier, our proposed DELIGHT detector can be integrated with the main detector's trigger system. This allows for correlating a significant missing transverse energy registered in the FCC-hh main detector in the direction of the DELIGHT detector with the LLP background at DELIGHT from energetic neutrinos. By leveraging this capability, events associated with these neutrinos can be effectively vetoed. We plan to explore the feasibility and effectiveness of such a veto in a future study.}\\

\textbf{\textit{Shielding $-$}} In the FCC-hh baseline design, the combined ECAL+HCAL system corresponds to approximately 10.5 nuclear interaction lengths ($\lambda_I$)\,\cite[Chapter 8]{FCC:2018vvp}, which is not sufficient to fully contain hadronic showers of the high energy particles produced at the IP.
\begin{figure}[hbt!]
    \centering        
    \includegraphics[width=0.6\textwidth]{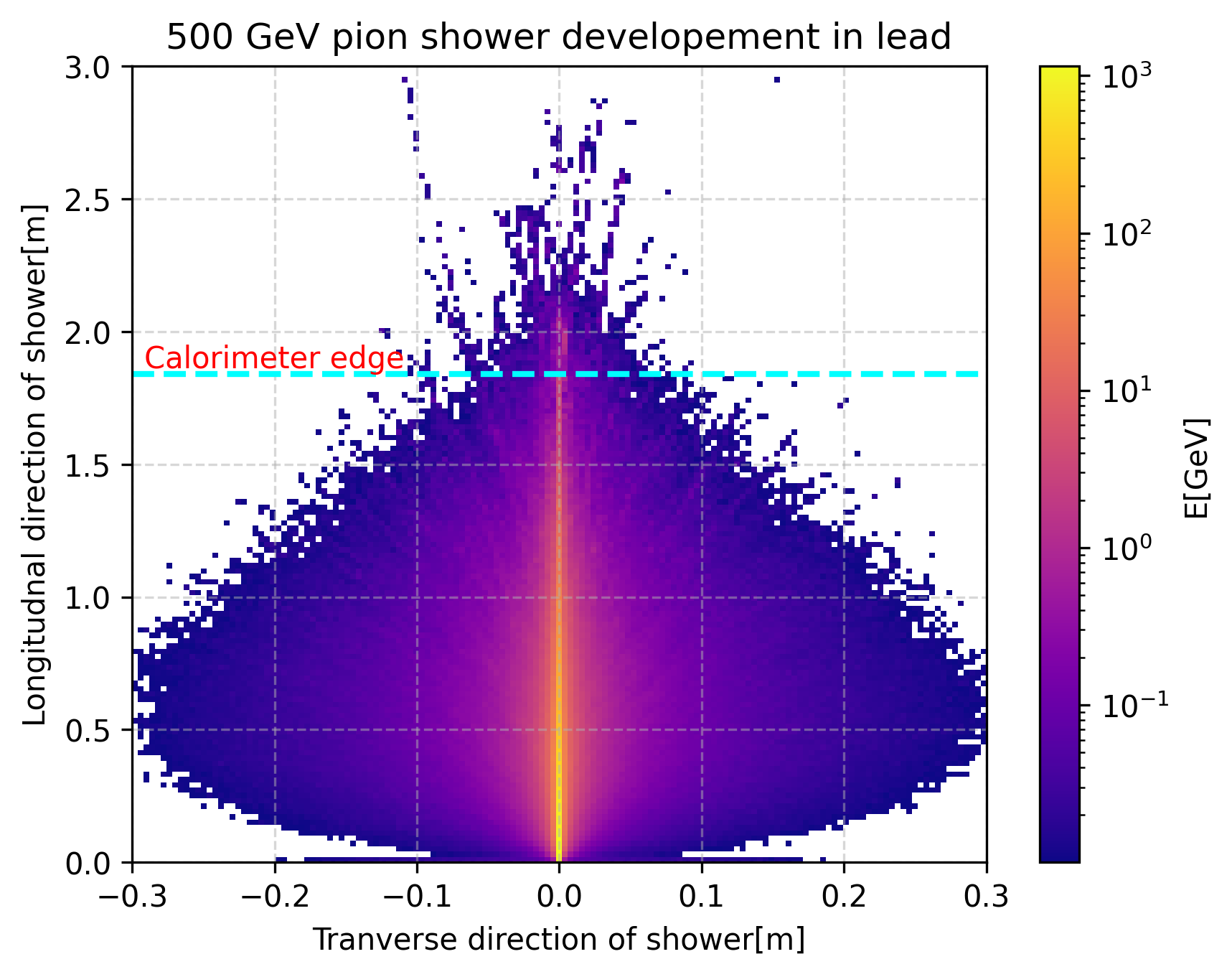}
    \caption{Development of a hadronic shower initiated by a 500\,GeV charged pion in a lead block of thickness $10.5\,\lambda_I$ (corresponding to 1.84\,m), simulated using \textsc{Geant4}. The plot shows the longitudinal and transverse energy deposition profile of the shower, averaged over $10^4$ events. The depth of $10.5\,\lambda_I$ is chosen to approximately model the combined ECAL+HCAL response of the FCC-hh main detector. While the bulk of the shower is contained within the calorimeter volume, a non-negligible fraction of secondary particles is observed to leak beyond the downstream edge, motivating the need for additional shielding.}
    \label{pion_shower_in_main_det}
\end{figure}
To quantify this effect, we perform a dedicated \textsc{Geant4} simulation in which a lead block of thickness 10.5$\lambda_I$(corresponding to 1.84\,m) is used to emulate the response of the main detector calorimeters. A 500\,GeV charged pion was injected using the \textsc{Geant4} particle gun. Averaged over 10\,000 events, the longitudinal shower profile is shown in Fig.\,\ref{pion_shower_in_main_det}, which clearly shows that while the bulk of the shower is contained, a non-negligible fraction of secondary particles leaks out of the downstream face of the calorimeter. The dominant leaking secondaries include neutrons, protons, charged pions, and a small fraction of kaons and muons. FCC-hh is expected to produce such high energy particles abundantly, for example, around $10^8$ pions are produced with energies $\sim$500\,GeV from b$\bar{\text{b}}$ production alone.

%To assess the relevance of this effect at FCC-hh compared to HL-LHC, we generate high-$p_T$ pions from b$\bar{\text{b}}$ production using \textsc{Pythia8} (HardQCD with $\hat{p}_T$=1\,TeV), at both $\sqrt{s}=100$ and 14\,TeV, respectively. The resulting spectra were scaled to the expected integrated luminosities of FCC-hh and HL-LHC and evaluated within $|\eta|<$2.4. Resulting distributions are shown in Fig.\,\ref{pion_from_bb}, which clearly shows that the FCC-hh spectrum is significantly harder and more copious compared to HL-LHC. In particular, the plot shows that FCC-hh is expected to produce around $10^8$ pions with energies around 500\,GeV from b$\bar{\text{b}}$ production alone, compared to only $10^4$ such pions at the HL-LHC.

%Consequently, although the calorimeter interaction lengths are similar (10.5 for FCC-hh, 9-10 for HL-LHC), the substantially higher energies and fluxes of particles at FCC-hh lead to an increased probability of hadronic leakage toward a downstream LLP detector such as DELIGHT. This motivates the inclusion of dedicated shielding to suppress these backgrounds.

\begin{figure}[hbt!]
\centering
\includegraphics[width=0.55\textwidth]{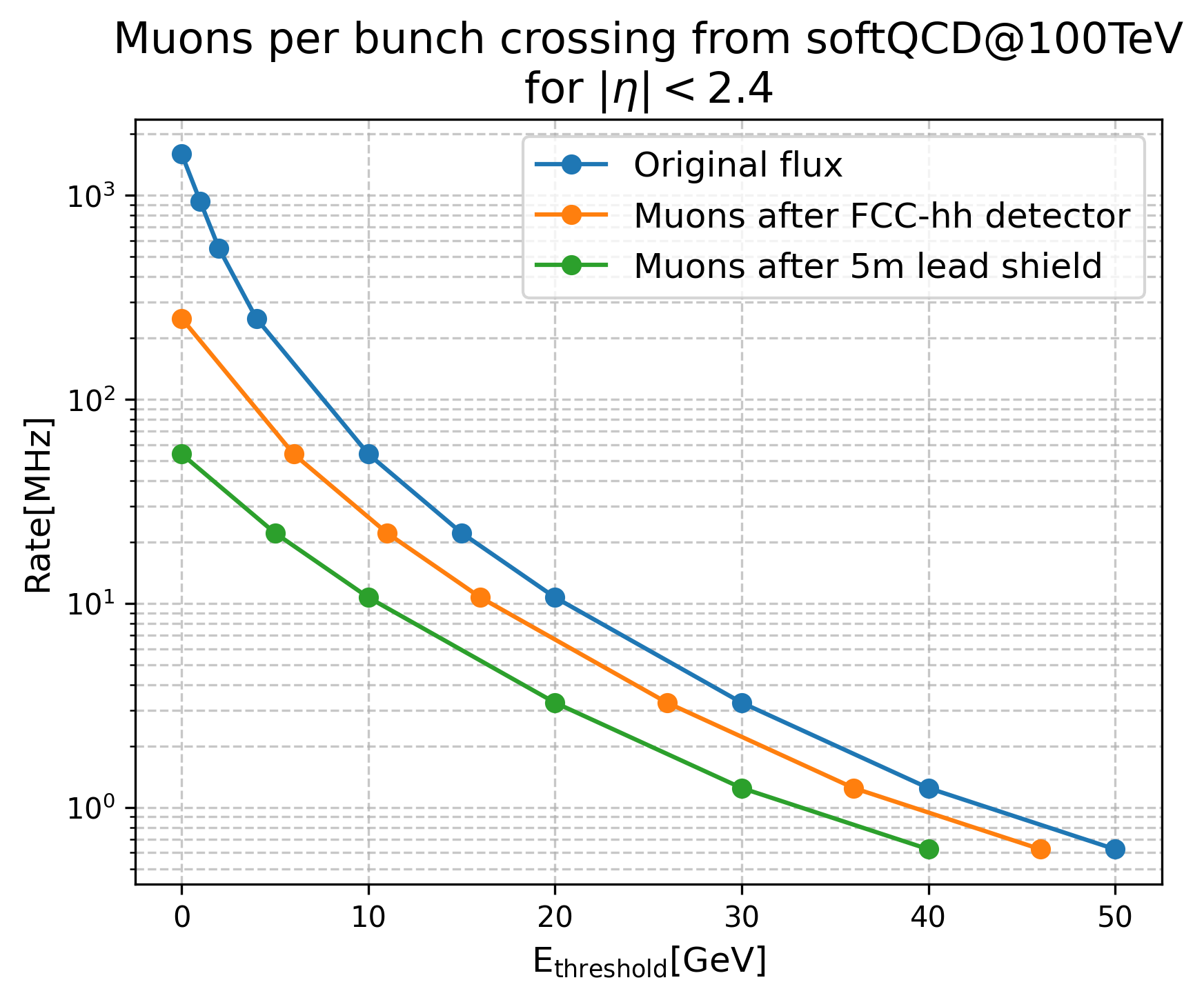}% plot
\caption{Muon rates from soft-QCD processes at the FCC-hh ($\sqrt{s}=100$~TeV) as a function of the muon energy threshold. The distributions are shown for the pseudorapidity range $|\eta|<2.4$.
The blue curve shows the initial muon flux produced from the interaction point, while the orange curve represents the spectrum after energy loss in the FCC-hh main detector, assuming an average energy loss of $\sim$3--4\,GeV for minimum-ionizing muons. The green curve shows the further attenuation of the muon flux after an additional 5\,m thick lead shielding placed downstream of the main detector, corresponding to an additional energy loss of $\sim$6--7\,GeV. The rate calculations include the effect of high pile-up (PU=1000)\,\cite{FCC:2018vvp,FCC:2025uan} conditions at FCC-hh. The figure illustrates that without additional shielding a large flux of penetrating muons would reach the LLP detector, motivating the need for dedicated shielding to suppress this background.}
\label{muon_from_softQCD} 
\end{figure}

An additional motivation for dedicated shielding before the  DELIGHT detector is the suppression of the muon background. 
%The {\it blue} curve in Fig.\,\ref{muon_from_softQCD} shows the muon flux originating from soft-QCD processes at FCC-hh with a pile-up of 1000, as a function of the muon energy threshold.
%The muon flux reaching the DELIGHT detector at the FCC-hh would be very high.
A minimum-ionizing muon is expected to lose approximately 3–4\,GeV of energy while traversing the FCC-hh main detector\,\cite[Chapter 8]{FCC:2018vvp}. Consequently, muons with energies above this threshold will penetrate the full detector volume and the muon energy spectrum downstream of the main detector will effectively shift by about 4\,GeV, as can be seen in Fig.\,\ref{muon_from_softQCD}.
%, as illustrated by the {\it orange} curve in Fig.\,\ref{muon_from_softQCD}. For instance, muons produced with an initial energy of 5\,GeV, corresponding to a rate of roughly 200 MHz, would emerge from the main detector with energies of about 1\,GeV at a similar rate.

If left unsuppressed, such a large flux of penetrating muons would overwhelm the DELIGHT detector and severely compromise its performance for LLP searches by coinciding with the signal and getting the event vetoed. To mitigate this background, additional downstream shielding is therefore required. As an illustrative example, the inclusion of a 5\,m thick lead shield following the main detector would absorb approximately another 6–7\,GeV of energy from minimum-ionizing muons. This results in a cumulative energy loss of about 10\,GeV, effectively shifting the muon spectrum further to lower energies.
%, as shown by the {\it green} curve in Fig.\,\ref{muon_from_softQCD}.
With this additional shielding in place, the flux of 10\,GeV muons is reduced from approximately 50\,MHz to about 10\,MHz.
%, corresponding to roughly one such muon per four bunch crossings (assuming a 40\,MHz collision rate). This demonstrates that dedicated shielding is essential to suppress the otherwise overwhelming muon background and to ensure the feasibility of DELIGHT for LLP studies. In order to study the structural issues introduced due to this shielding, one would require to do a dedicated study which is beyond the scope of the present work.}
For FCC-ee, where the estimated background rates are much lower, a shielding of even 2-3\,m of lead is enough to block all the SM backgrounds.

We now shift our focus to the forward production of LLPs.

\section{Forward physics at FOREHUNT}
\label{sec:forward}
\vspace*{-0.2cm}

To demonstrate the potential of dedicated detectors to explore forward new physics, we study the production of LLPs $\phi$ from $B$-meson decays.
A forward dedicated detector, FASER (ForwArd Search ExpeRiment)~\cite{Feng:2017uoz,FASER:2018eoc,FASER:2019aik}, is already designed to look for light LLPs at the LHC in the far forward region, placed 480\,m away from the ATLAS IP.
%\sout{It is expected to collect 150\,fb$^{-1}$ of data in Run-3 of LHC.} \AS{
With LHC Run 3 nearing completion in 2026, the FASER experiment is on track to surpass its initial data collection goals, with an expected total dataset of roughly $250\text{ fb}^{-1}$ by the start of Long Shutdown 3\,\cite{abraham2024neutrino}, and will continue collecting data in Run 4.
At the HL-LHC, an upgrade called FASER 2\,\cite{Feng:2022inv} is proposed with a much bigger detector volume designed to collect 3\,ab$^{-1}$ of data.
At 100\,TeV, $B$-mesons have an increased forward production with higher momenta compared to 14\,TeV, significantly boosting the detection of LLPs in a forward detector. 
This enhancement is prominent for LLPs with small $c\tau$ values and higher masses, as shown in Ref.\,\cite{Bhattacherjee:2023plj}. 
For example, just changing the center-of-mass energy from 14\,TeV to 100\,TeV for the FASER 2 detector, the signal acceptance increases by a factor of 30 for $m_\phi=0.1$\,GeV and $c\tau=10^{-2}$\,m.
For $m_{\phi}=4.4$\,GeV having the same decay length, the signal acceptance still increases by a considerable factor.
%of $10^{21}$.
\begin{figure}[htb!]
    \centering
    \includegraphics[width=0.46\textwidth]{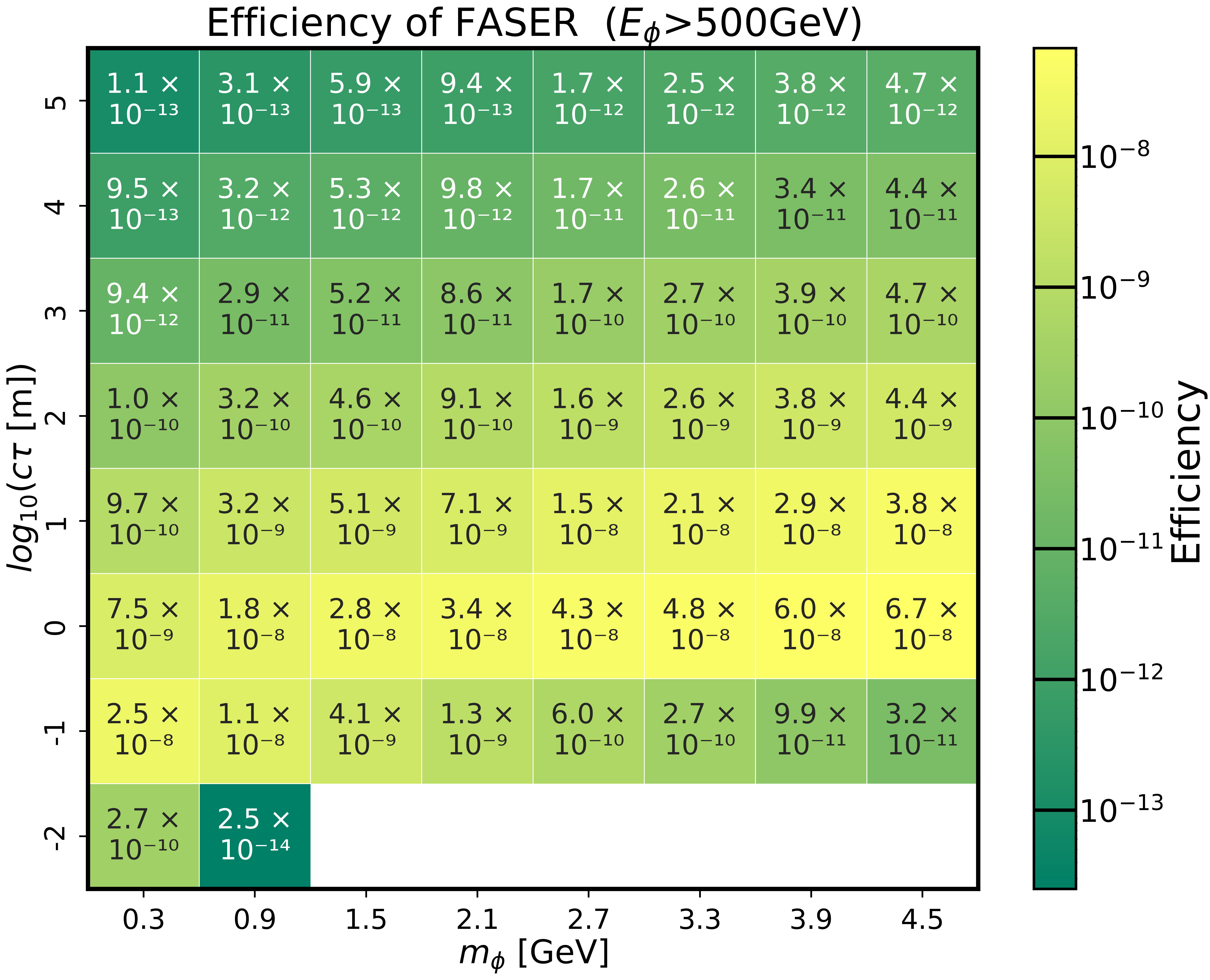}\hspace{1cm}
    \includegraphics[width=0.46\textwidth]{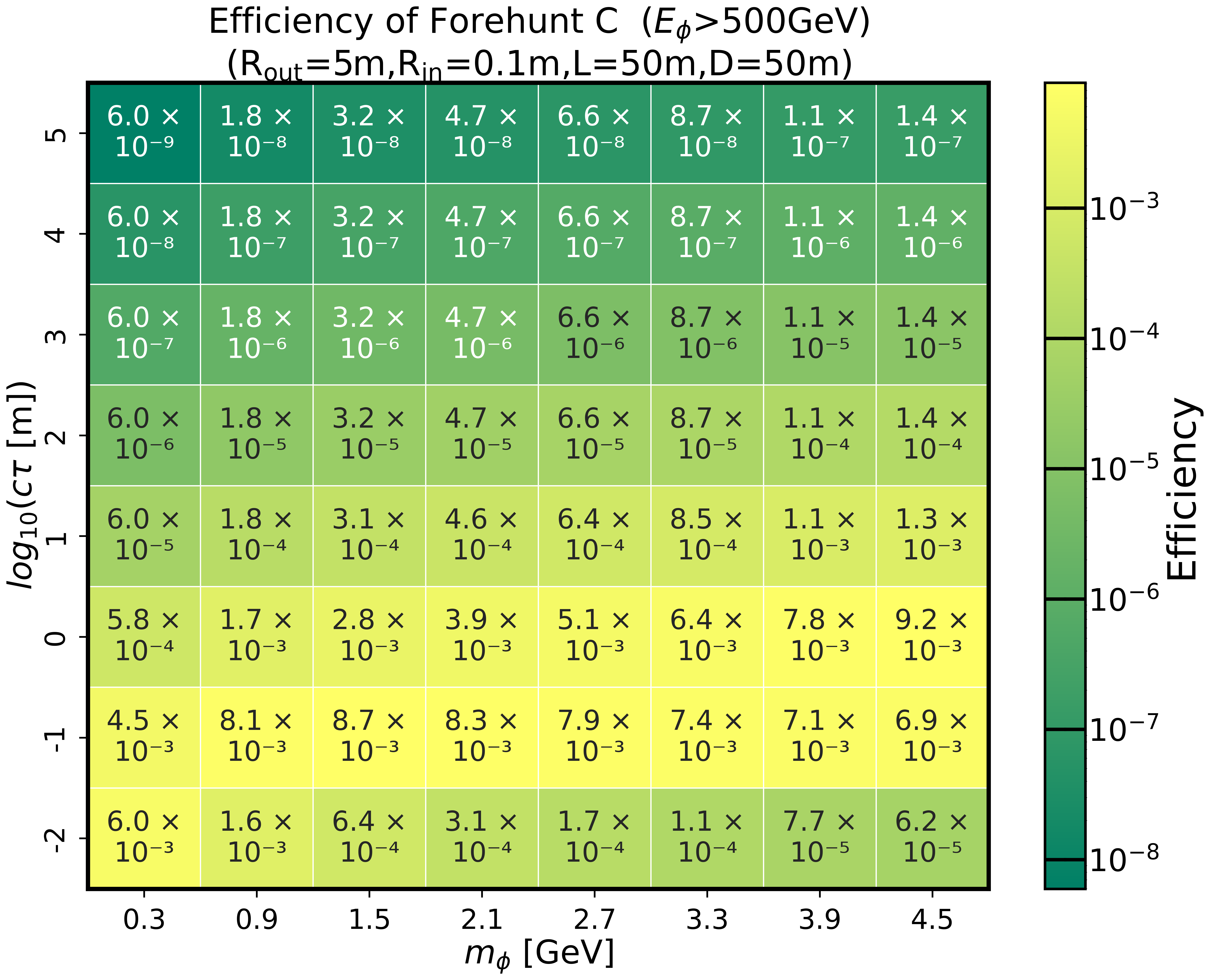}
    \caption{Detection efficiencies of forward LLP detectors for the dark Higgs model in the $(m_\phi, c\tau)$ plane, assuming LLP production from $B$-meson decays. The {\it left} panel shows the efficiency for the FASER detector at $\sqrt{s}=14$~TeV, while the {\it right} panel shows the efficiency for the FOREHUNT-C configuration at $\sqrt{s}=100$~TeV. In both cases, only LLPs with energy $E_\phi > 500$\,GeV are considered to ensure detectability. For FOREHUNT-C, a cylindrical region along the beam axis is excluded to account for the beam pipe. The efficiencies represent the probability of LLP to decay within the detector volume. The FOREHUNT-C configuration, shown here for illustration, corresponds to an idealized geometry with large radius and proximity to the interaction point, providing an upper benchmark on the achievable sensitivity in the forward direction.}
    \label{fig:FASER_FOREHUNT}
\end{figure}

However, we can enhance the sensitivity even further by optimizing the detector size and its placement.
We observe that with increasing distance from the IP, the signal acceptance falls rapidly. Increasing the detector's length and radius further enhances the signal acceptance.
For the 100\,TeV FCC-hh, we proposed the FOREHUNT detector~\cite{Bhattacherjee:2023plj} ({\bf For}ward {\bf e}xperiment for {\bf hun}dred {\bf T}eV).
The FCC-hh main detector is proposed to be placed at $z\in[-25,25]$\,m~\cite{Mangano:2017tke,Mangano:2022ukr,FCC:2018vvp,FCC:2025lpp}. 
With a buffer length of 25\,m, we put forward the proposal to place FOREHUNT at a minimum distance of 50\,m from the FCC-hh IP. 
Among the six configurations considered in Ref.\,\cite{Bhattacherjee:2023plj}, FOREHUNT-C provides the best sensitivity due to its proximity to the IP and larger radius and length. 
Additionally, we discuss a proposed design of the FOREHUNT detector in Ref.\,\cite{Bhattacherjee:2023plj}, with a shielding block and scintillator layers both before and after the shielding block to control the backgrounds from muons, neutral hadrons, and neutrinos.

We analyze the detection efficiency of FASER at 14\,TeV and FOREHUNT-C at 100\,TeV for dark Higgs bosons produced via $B$-meson decays.
We require the energy of the LLPs reaching the forward detectors to be greater than 0.5\,TeV to ensure their detection. 
We also remove a narrow cylindrical region of 10\,cm radius from the centre to account for the beam pipe, as illustrated in Fig.\,\ref{fig:FCC_FOREHUNT}.
The detection efficiencies in the ($m_\phi$, $c\tau$) plane are shown in Fig.\,\ref{fig:FASER_FOREHUNT}, which demonstrates that FOREHUNT-C ({\it bottom} panel) provides a significant improvement compared to FASER ({\it top} panel)\,\footnote{The performance of FOREHUNT-C for LLPs produced from $B$-meson decays at the FCC-ee is studied in Ref.\,\cite{Bhattacherjee:2025dlu}}. 
We do not show the FASER efficiency for mass and lifetime points where the product of the detector's efficiency and total $B$-meson production during the HL-LHC run is less than one, as this would result in fewer than one expected detected event, even with a branching ratio of one for the decay $B\rightarrow K\phi$.

Note that the DELIGHT detector also provides sensitivity for light LLPs coming from $B$-meson decays, as shown in Ref.\,\cite{Bhattacherjee:2023plj}.
Moreover, the efficiencies of detecting LLPs per bunch crossing produced in decays of mesons would increase in the high pile-up environment of the FCC-hh, as pointed out in Ref.\,\cite{Bhattacherjee:2025pxg}.

%\section{Optimization of the dedicated detector concepts}
%\label{sec:optimise}
%DELIGHT-opt

\subsection{Optimization of the forward detector}
\vspace*{-0.2cm}

\begin{figure*}[htb!]
\centering
    \includegraphics[width=0.32\textwidth]{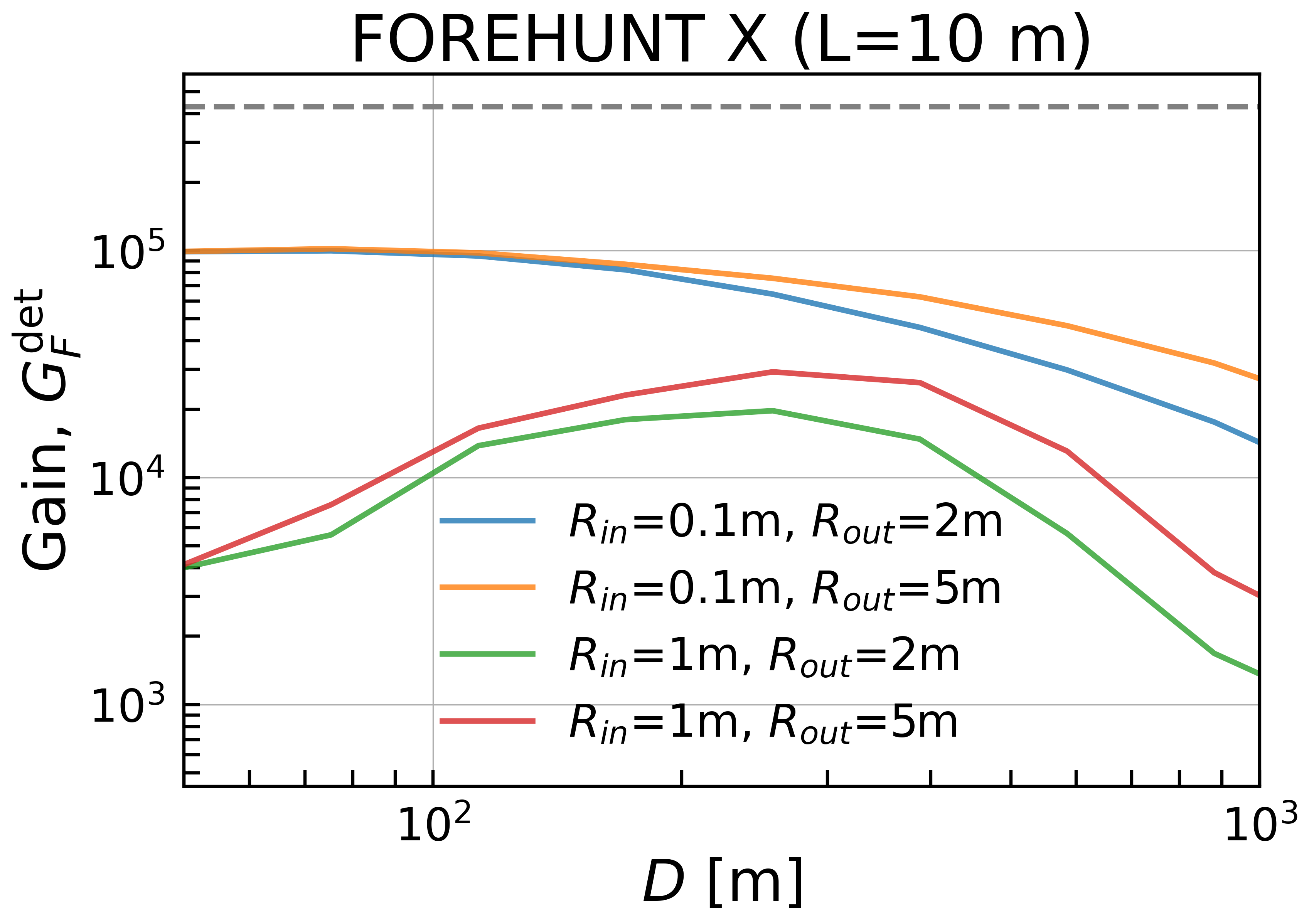}
    \includegraphics[width=0.32\textwidth]{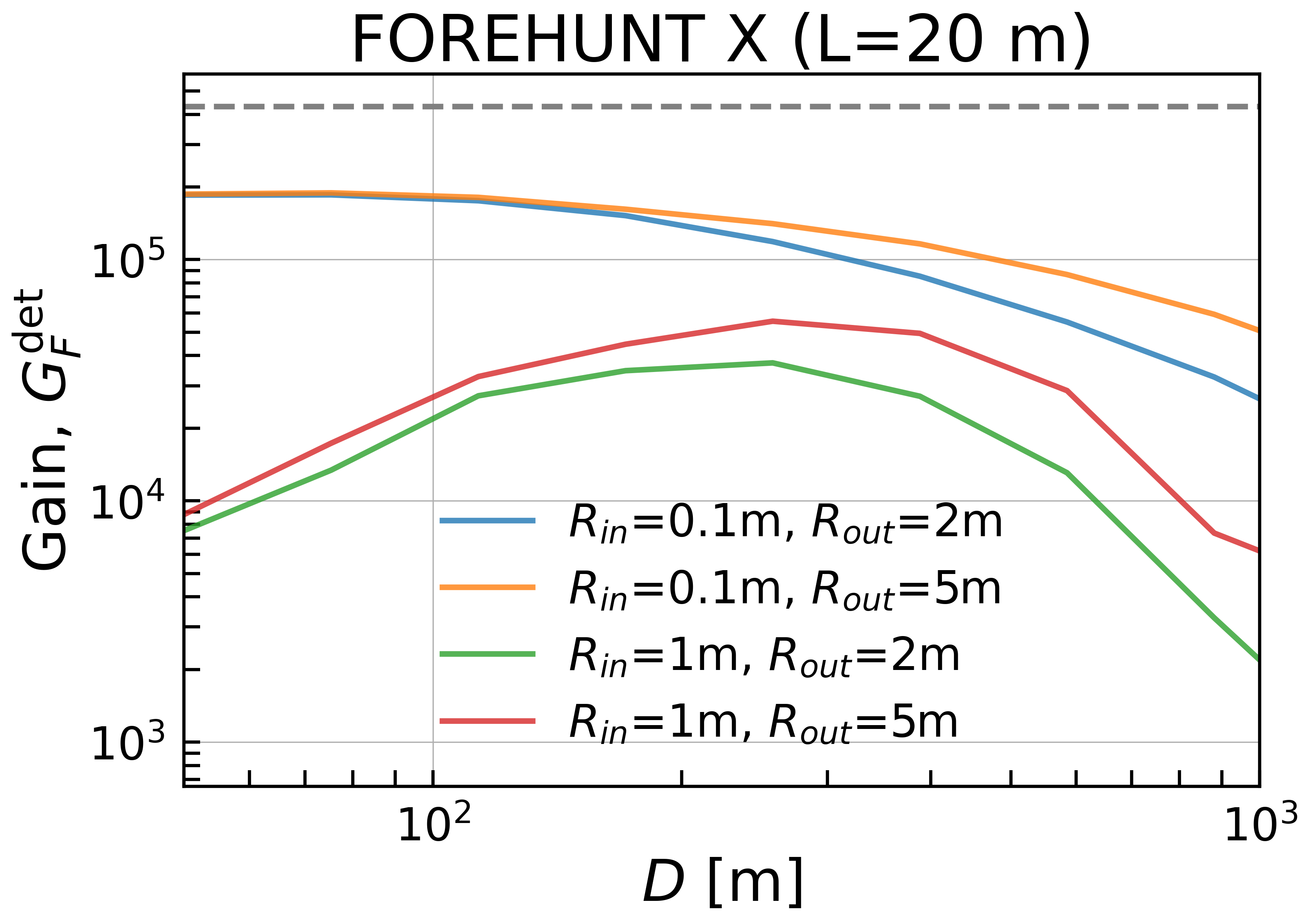}
    \includegraphics[width=0.32\textwidth]{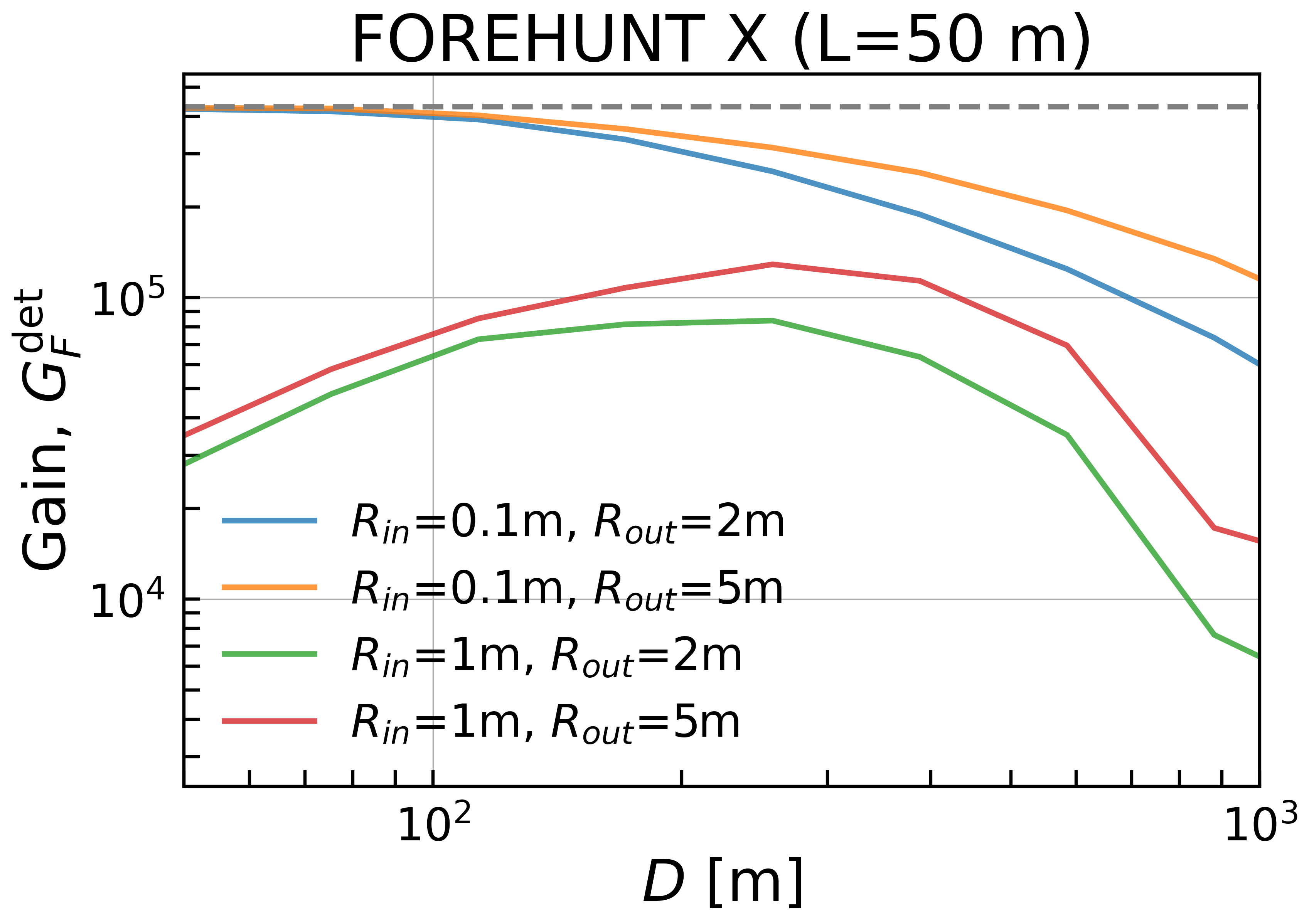}
    \caption{Gain in detection efficiency, $G_F^{\text{det}}$, relative to FASER, for forward LLP detectors as a function of the detector distance from the interaction point ($D$). Results are shown for different inner ($R_{\rm in}$) and outer ($R_{\rm out}$) radii, and for three detector lengths ($L = 10\,\mathrm{m},\,20\,\mathrm{m},\,50\,\mathrm{m}$). The study is performed for the dark Higgs model with LLP production via $B \rightarrow \phi K$ at $\sqrt{s}=100$~TeV. The gain is defined in Eq.~(\ref{eq:GF_det}) and is averaged over a representative range of LLP masses and lifetimes in the $(m_\phi, c\tau)$ parameter space, as illustrated in Fig.~\ref{fig:FASER_FOREHUNT}. A minimum LLP energy requirement of $E_\phi > 500$~GeV is imposed. The figure highlights the trade-off between detector size and placement in optimizing sensitivity. The dashed line corresponds to the gain obtained for the idealized FOREHUNT-C configuration, shown for reference.}
    \label{fig:forward_opt}
\end{figure*}

%FOREHUNT-opt
We now perform a similar optimization for detectors in the forward direction as we did for the transverse detector.
For a cylindrical forward detector along the beam axis, the parameters affecting the performance are the inner radius, $R_{\text{in}}$, outer radius, $R_{\text{out}}$, detector length, $L$, and the distance of the detector from the IP, $D$.
We vary $D$ from 50\,m to 1\,km, for three different detector lengths and two possible values of $R_{\rm in}$ and $R_{\rm out}$ each.
We consider all the 64 points in the ($m_\phi$, $c\tau$) plane shown in Fig.\,\ref{fig:FASER_FOREHUNT} as benchmarks in this optimization.
Here, we quantify the performance gain of each of these forward detectors, $G_F^{\rm det}$, over the FASER detector using
\begin{equation}
G_F^{\rm det} = \left [  \prod_{\rm benchmarks} \text{min}\left(\frac{\epsilon_{\rm det}}{\epsilon_{\rm FASER}} ,10^{10}\right) \right ]^{\textstyle \frac{1}{64}},
\label{eq:GF_det}
\end{equation}
where $\epsilon_{\rm det}$ and $\epsilon_{\rm FASER}$ are the efficiencies for each of the 64 benchmark points.
If, for any detector, the gain for a particular benchmark is more than $10^{10}$ compared to FASER, we have capped it at $10^{10}$.
We show the gain, $G_F^{\rm det}$, as a function of $D$ for varying values of $L$, $R_{\rm in}$ and $R_{\rm out}$ in Fig.\,\ref{fig:forward_opt}.
The FOREHUNT-C proposed by us has an average gain of approximately $4\times10^5$.

We observe that FOREHUNT-C consistently outperforms all the alternative configurations.
For configurations with a small $R_{\text{in}}$, the detection efficiency decreases significantly with increasing detector distance, being 5–6 times lower at 1\,km.
Configurations with a larger $R_{\text{in}}$ initially exhibit an increase in detection efficiency as the detector distance from the IP increases, followed by a subsequent decline. 
This behavior arises due to a large cylindrical region in the forward direction devoid of detection material, which accommodates the beam pipe.  
When the detector is positioned very close to the IP, a significant portion of the forward particle flux escapes through this inner cylindrical region, as the $\eta$ range of this region is high. 
However, as the detector is moved further away, the $\eta$ range of the inner cylinder for a fixed $R_{\rm in}$ reduces, restoring a portion of the forward flux and leading to an increase in efficiency.
%particles reach a sufficient radial distance to enter the annular detection region, leading to an increase in efficiency.}
These configurations reach peak efficiency at approximately 200--300\,m from the IP, with the maximum efficiency being nearly an order of magnitude greater than at a 1\,km placement. It is evident from Fig.\,5 that even at the distance with the peak efficiency, FOREHUNT X with $L=50~\text{m},~R_{\text{in}}=1~\text{m},~\text{and}~R_{\text{out}}=5~\text{m}$ is approximately 4 times less efficient than FOREHUNT-C.
Beyond this optimal distance, the efficiency decreases primarily due to the increasing $D$, which reduces the solid angle coverage and affects the detection probability. 

\subsection{The hybrid-FOREHUNT}
\vspace*{-0.2cm}

\begin{figure}[hbt!]
    \centering
\includegraphics[width=0.7\textwidth]{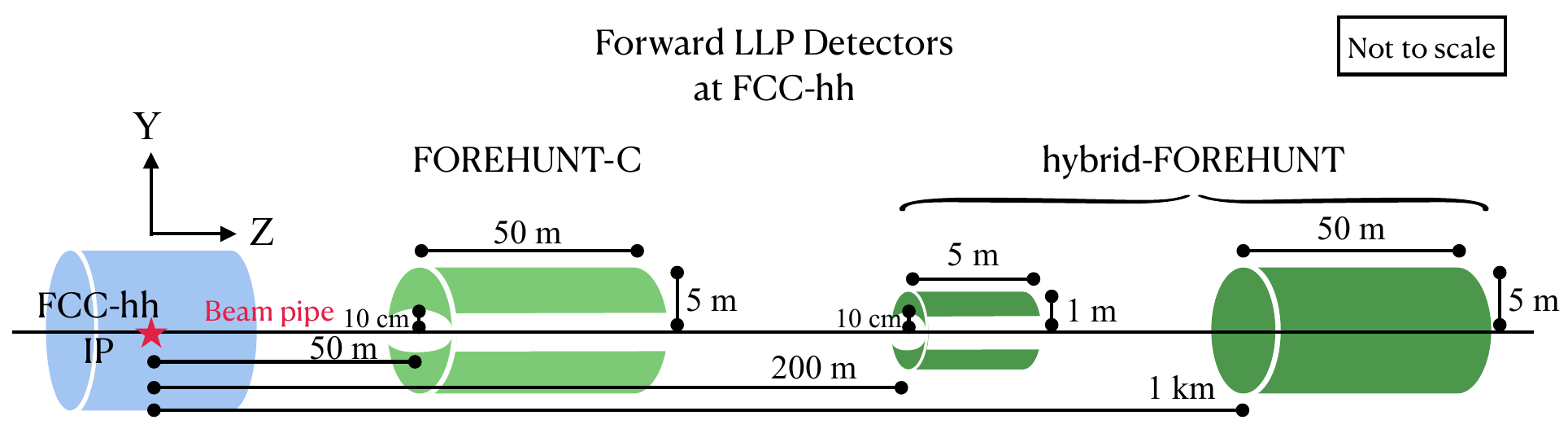}
    \caption{Schematic illustration of the forward detector geometries considered in this work, including the FOREHUNT-C and hybrid-FOREHUNT configurations. The detector is modeled as a cylindrical volume characterized by its distance from the interaction point ($D$), length ($L$), inner radius ($R_{\rm in}$), and outer radius ($R_{\rm out}$). A cylindrical region along the beam axis is excluded to account for the beam pipe and beamline infrastructure, which is particularly relevant for detectors placed close to the interaction point. The FOREHUNT-C setup, shown here, corresponds to an idealized configuration with minimal inner radius and close proximity to the interaction point, and serves as an upper benchmark on achievable sensitivity. The hybrid-FOREHUNT configuration combines a near detector and a far detector, designed to retain sensitivity to short-lived LLPs while extending coverage to longer lifetimes.}
    \label{fig:FCC_FOREHUNT}
\end{figure}

\begin{figure}[hbt!]
   \centering    
\includegraphics[width=0.5\textwidth]{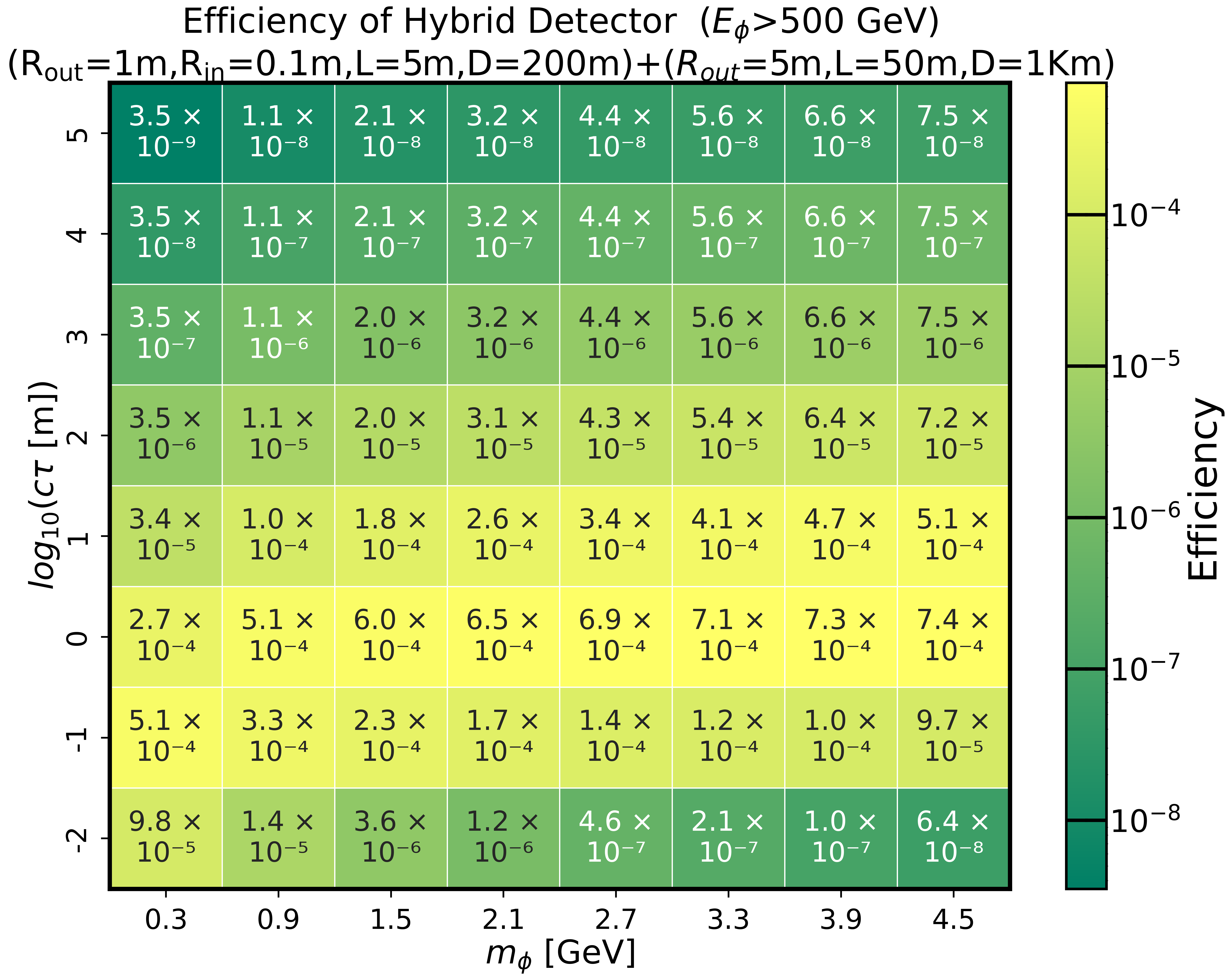}
   \caption{Detection efficiency of the hybrid forward detector configuration in the $(m_\phi, c\tau)$ plane for the dark Higgs model, assuming LLP production via $B \rightarrow \phi K$ at $\sqrt{s}=100$~TeV and  requiring $E_\phi > 500$~GeV. The hybrid setup consists of a near detector located at $200~\mathrm{m}$ from the interaction point and a far detector placed at 1\,$\mathrm{km}$, with geometrical parameters (including inner radius $R_{\rm in}=10\,\mathrm{cm}$ for beam pipe) indicated in the plot. The efficiency shown corresponds to the probability that a LLP decays within either of the two detector volumes. The hybrid configuration combines the sensitivity of the near detector to shorter lifetimes with the extended reach of the far detector for long-lived scenarios, resulting in improved coverage over a broad region of parameter space.}
  \label{fig:forward_near_far}
\end{figure}

Our analysis indicates that, for optimal detection efficiency, the detector should be positioned as close to the IP as possible. While FOREHUNT-C yields the highest sensitivity, its proximity to the IP might lead to practical integration constraints; consequently, we investigated alternative FOREHUNT-X geometries. However, these configurations still necessitate the placement of a large-scale detector in close proximity to the IP, which may remain technically or logistically prohibitive.
In case such a placement is not feasible, alternative configurations must be considered to minimize efficiency loss. 
We find that a hybrid setup, combining a smaller detector positioned near the IP with a larger detector placed at a greater distance, can yield comparable performance. 
In this arrangement, the near detector compensates for the reduced efficiency of the far detector in regions corresponding to small particle decay lengths. 
Also, this hybrid configuration is expected to face lower SM backgrounds as compared to FOREHUNT-C or FOREHUNT-X.
The far detector can be placed approximately 1\,km away from the FCC-hh IP. 
The corresponding efficiencies for this combined detector configuration, hybrid-FOREHUNT, are presented in Fig.\,\ref{fig:forward_near_far}. We also show the individual contributions of the near and far detectors in Figs.\,\ref{fig:forward_near} and \ref{fig:forward_far} in the Appendix.
For comparison, the hybrid-FOREHUNT yields a gain four times lower than that of FOREHUNT-C. However, it consistently outperforms FOREHUNT X with $L=10~\text{m}$ and is comparable to FOREHUNT X with $L=20~\text{m}$ and $L=50~\text{m}$.
The configuration of hybrid-FOREHUNT is also shown in Fig.\,\ref{fig:FCC_FOREHUNT}. The FCC-hh beam pipe bends around 1.25\,km from the IP\,\cite{Benedikt:2016pc}.
Therefore, the placement of far detector at 1\,km may not be possible due to beam bending constraint, we computed similar efficiency plots after placing the far detector at a distance of 2\,km, which is shown in Fig.\,\ref{fig:hybrid_2km} in the Appendix.  

For the detector configurations with the beam pipe traversing the detector volume, one would expect substantial beam-induced backgrounds and radiation effects. A quantitative assessment of these effects would require a dedicated background and radiation study, which is beyond the scope of the present work. Nevertheless, related studies exist in the literature, for example in the context of forward detectors at the HL-LHC\,\cite{Hacisahinoglu:2025xrs}.
At a qualitative level, several mitigation strategies could be envisaged, including shielding the inner region of the detector with dense material to absorb beam-induced backgrounds, incorporating tracking capabilities to veto backgrounds that can be extrapolated back to the interaction point, or a combination of both. Such measures would inevitably increase the effective inner radius of the decay volume and correspondingly reduce the sensitivity. A detailed evaluation of these effects is, therefore, left for future dedicated studies.

In a nutshell, we demonstrate the role of the DELIGHT and FOREHUNT proposals 
in LLP searches.
They fit in case of either of the scenarios discussed in Section\,\ref{sec:role}.
Even if we observe any hint of displaced new physics in one of the currently approved experiments, like SHiP, these future dedicated detectors would be crucial in investigating the nature of the new physics, for example, by being sensitive to a complementary production mode not accessible to the previous experiments.
The closeness of the detectors to the FCC-ee/FCC-hh IPs opens up the possibility of integrating them to the main detector trigger systems.
We have also identified the core-DELIGHT and hybrid-FOREHUNT 
options that might be more feasible to construct, while still maintaining some gain over the existing proposed detectors.

\section{Conclusion}
\label{sec:concl}
\vspace*{-0.2cm}

In this work, we present and justify the concept of an optimized and shared transverse detector design for both FCC-ee and FCC-hh.
%, alongside a forward detector for FCC-hh.
We modify our original proposal for a transverse detector at the FCC complex, named DELIGHT, to adapt it into a shared detector concept suitable for both the lepton and hadron colliders. 
Not only is this a sustainable solution, but it also significantly enhances sensitivity to light scalar LLPs from Higgs boson decays at both FCC-ee and FCC-hh.
Furthermore, we find that reducing the size of the detector or moving it farther from the IP beyond a certain point would not offer any additional benefit over general-purpose detectors, such as IDEA at FCC-ee. 
This leads us to our core-DELIGHT detector design.
%For FCC-hh, we propose the FOREHUNT detector to capture forward physics. 
%Our design prioritizes optimal physics performance by keeping the detector closer to the IP and increasing its size. 
%However, if this setup proves unfeasible, we present a hybrid option that combines smaller and larger forward detectors placed near and farther from the FCC-hh IP, respectively. 
%This hybrid approach would still preserve the performance of the original FOREHUNT design.
We believe that 
%both of 
this proposed detector would enhance sensitivity to other BSM scenarios that lead to displaced signatures.

Our study underscores that in order to maximize the potential of dedicated LLP detectors, they have to be kept at locations that are optimal for our physics goals.
Even if funding is not approved at this juncture, it is essential that CERN consider allocating space for these proposed detectors at their optimal locations. History has shown that unforeseen opportunities can arise, and being prepared is crucial. When the LHC commenced operations, the community anticipated the discovery of new high-mass particles around the corner. However, as the years passed without a discovery, researchers sought alternative locations within the accelerator and detector complex to accommodate detectors for LLPs. Unfortunately, this resulted in suboptimal placement and sizing for these detectors. In contrast, for the FCC-hh, we should prioritize foresight and flexibility. Proposal with significant physics potential, such as DELIGHT 
%and FOREHUNT, 
should not be dismissed solely due to temporary funding constraints. Instead, CERN should ensure that the required space for experimental caverns is reserved and prepared for future construction, allowing us to capitalize on emerging opportunities when funding becomes available.\\
%\clearpage

\textbf{\textit{Acknowledgements:}} B. Bhattacherjee acknowledges the MATRICS Grant (MTR/2022/000264) of the Science and Engineering Research Board (SERB), Government of India. The work of B. Bhattacherjee is also supported by the Core Research Grant CRG/2022/001922 of the Science and Engineering Research Board (SERB), Government of India. B. Bhattacherjee, C. Bose, and A. Sharma are grateful to the Center for High Energy Physics, Indian Institute of Science, for the cluster facility.
The work of N. Ghosh was supported by the Japan Society for the Promotion of Science (JSPS) as a part of the JSPS Postdoctoral Program (Standard), grant number: JP24KF0189, and by the World Premier International Research Center Initiative (WPI), MEXT, Japan (Kavli IPMU). 
S. Matsumoto is supported by a Grant-in-Aid for Scientific Research from the Ministry of Education, Culture, Sports, Science, and Technology (MEXT), Japan: 23K20232 (20H01895), 24H00244 (20H00153), and 24H02244 by the JSPS Core-to-Core Program: JPJSCCA20200002 and by the World Premier International Research Center Initiative (WPI), MEXT, Japan (Kavli IPMU).
The work of S. Mukherjee is supported by an initiation grant \texttt{(IITK/PHY/2023282)} received from IIT Kanpur. 

%\clearpage

\bibliographystyle{utphys.bst}
\bibliography{refs}

\clearpage

%\vspace*{1cm}

\appendix

\section*{Modifications and individual contributions to the hybrid-FOREHUNT}

\begin{figure}[hbt!]
   \centering    
\includegraphics[width=0.5\textwidth]{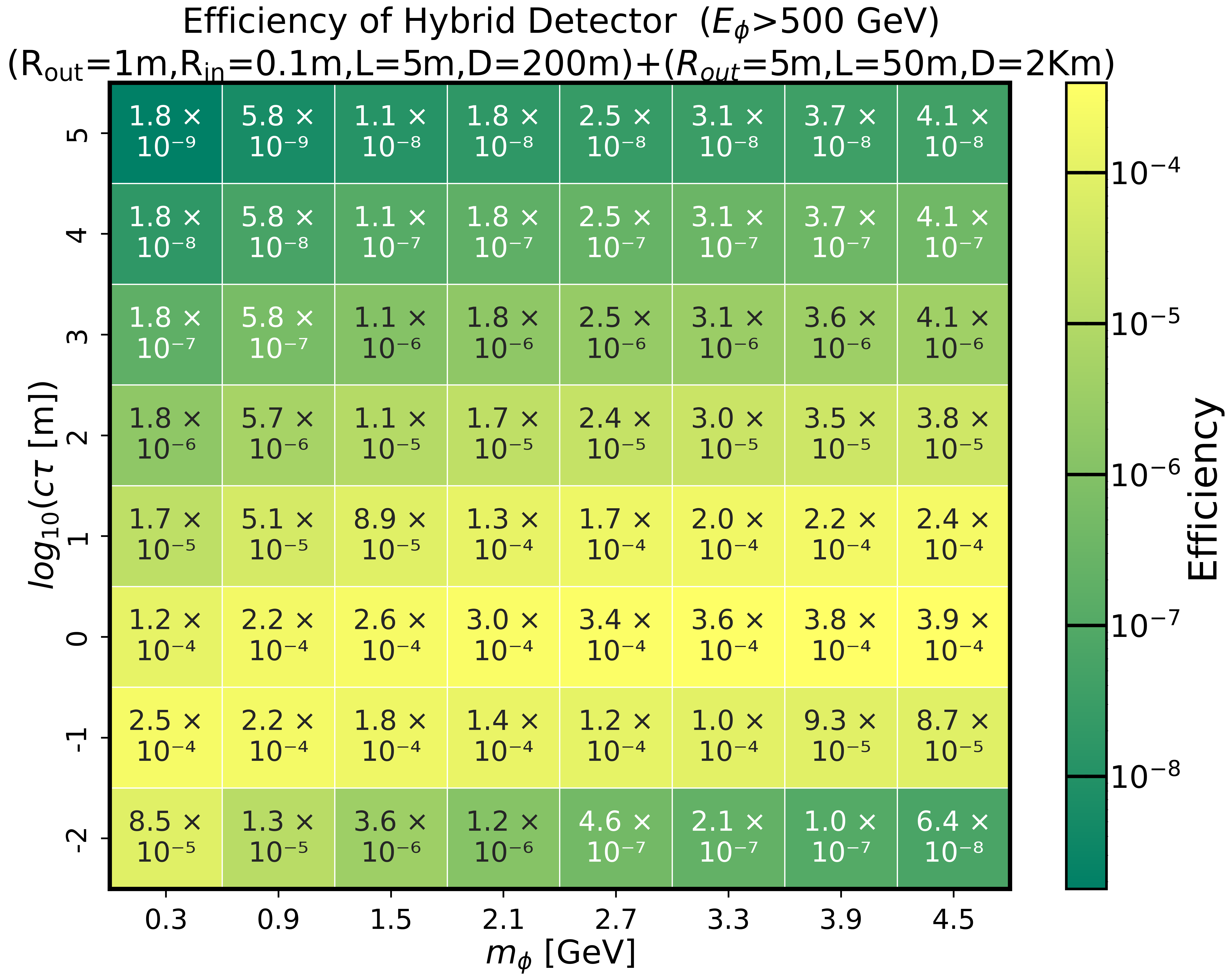}
   \caption{Detection efficiencies of the hybrid detector configuration for the dark Higgs model with $E_{\phi} > 500$ GeV. The parameters of the near and far detectors are mentioned in the title.}
  \label{fig:hybrid_2km}
\end{figure}

We show the efficiency grid for the hybrid-FOREHUNT, when the far detector is placed at a distance of 2\,km from the IP in Fig.\,\ref{fig:hybrid_2km}. We observe that the performance does not degrade much as compared to the far detector placed at 1\,km in Fig.\,\ref{fig:forward_near_far}.

\begin{figure}[hbt!]
   \centering    
\includegraphics[width=0.5\textwidth]{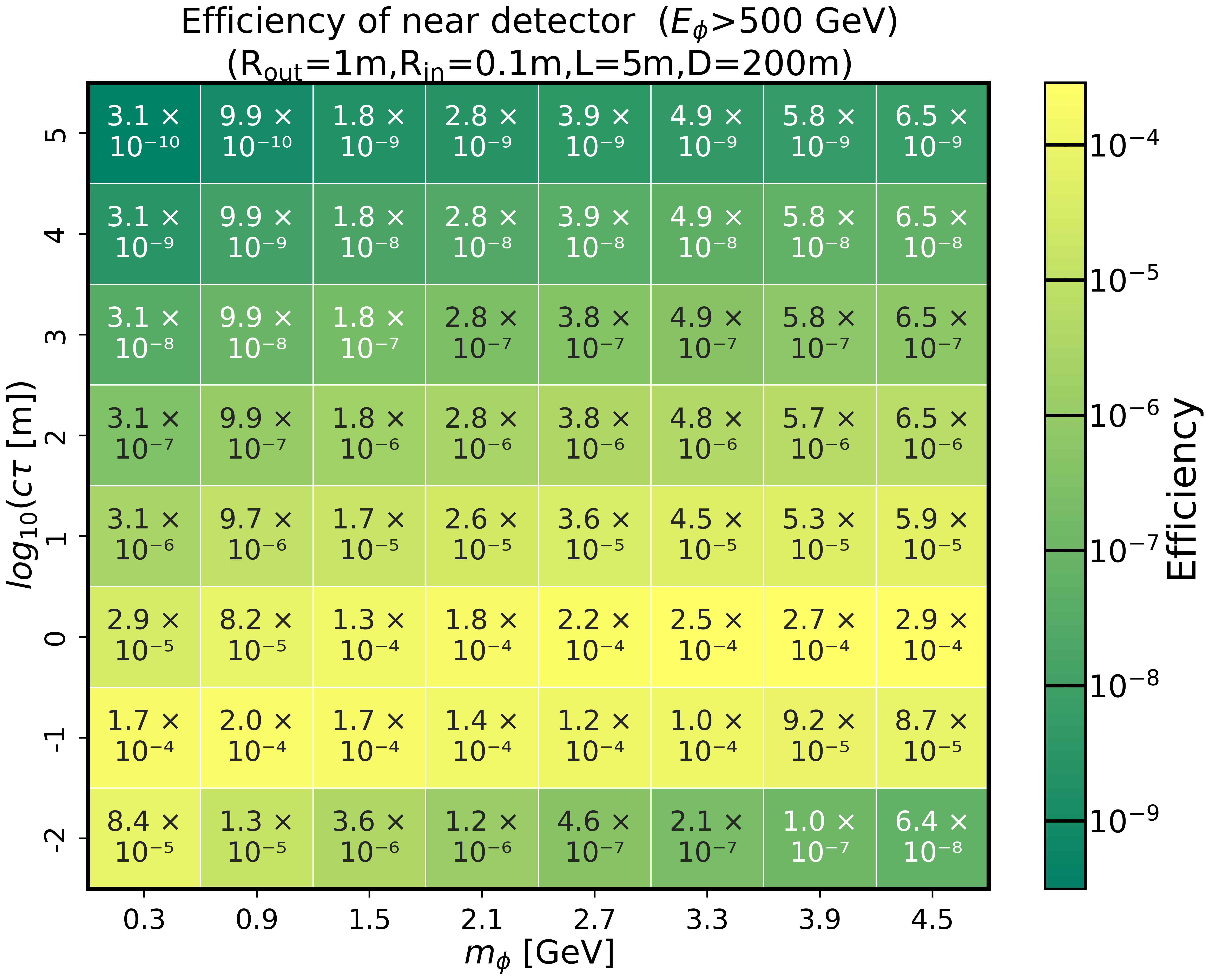}
   \caption{Detection efficiencies of the smaller and nearer detector part of the hybrid-FOREHUNT configuration for the dark Higgs model with $E_{\phi} > 500$\,GeV. The parameters of the near detector is mentioned in the title.}
  \label{fig:forward_near}
\end{figure}

\begin{figure}[hbt!]
   \centering    
\includegraphics[width=0.5\textwidth]{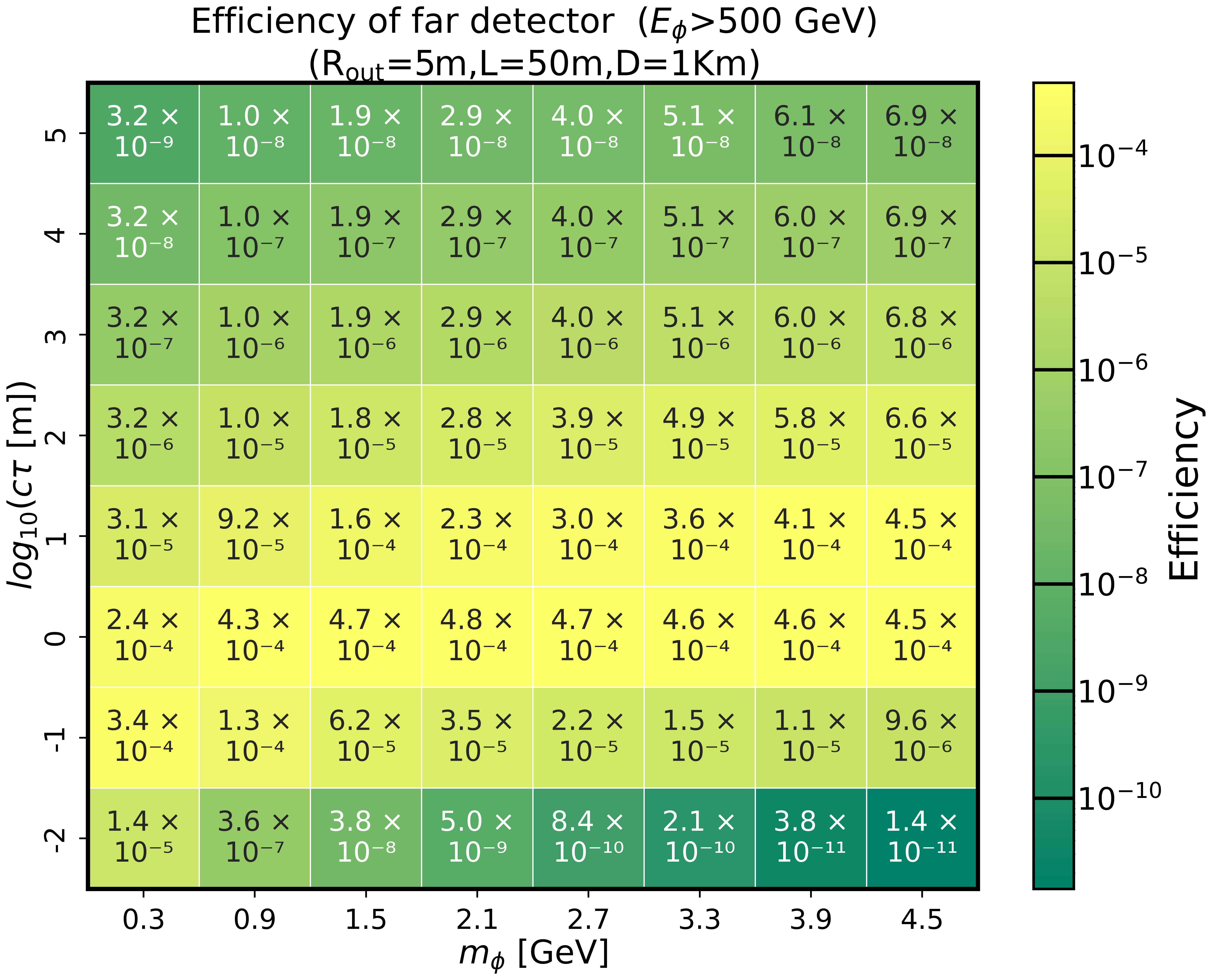}
   \caption{Detection efficiencies of the bigger and farther detector part of the hybrid-FOREHUNT configuration for the dark Higgs model with $E_{\phi} > 500$\,GeV. The parameters of the far detector is mentioned in the title.}
  \label{fig:forward_far}
\end{figure}

We also show the individual signal efficiencies of the near and far detectors of the hybrid-FOREHUNT configuration in Figs.\,\ref{fig:forward_near} and \ref{fig:forward_far}, respectively.

\end{document}